\documentclass[a4paper,fleqn,usenatbib,useAMS]{mnras}
\usepackage{cite, natbib}
\usepackage{graphicx, subfigure}
\usepackage{color}
\usepackage{bm}
\usepackage{amsmath}

\usepackage[T1]{fontenc}
\usepackage{ae,aecompl}
\usepackage{newtxtext,newtxmath}

\newcommand{\response}[1]{{#1}}
\newcommand{\Kepler}{{\it Kepler}}
\newcommand{\kepler}{\Kepler}

\newcommand{\LSST}{{\it LSST}}

\newcommand{\eg}{{\it e.g.}}

\newcommand{\nkoi}{1102}

\newcommand{\naigrain}{333}

\newcommand{\nkoimcq}{275}
\newcommand{\kepexample}{5809890}
\newcommand{\kepexampleperiod}{30.5}
\newcommand{\aigrainexampleperiod}{20.8}

\newcommand{\oldpgramRMS}{18.98}

\newcommand{\oldpercentpgramMAD}{4.70}

\newcommand{\pgramRMS}{1.04}
\newcommand{\pgramMAD}{0.52}

\newcommand{\percentpgramMAD}{2.82}

\newcommand{\gpRMS}{0.25}
\newcommand{\gpMAD}{0.40}

\newcommand{\percentgpMAD}{1.97}

\newcommand{\gpRMSnp}{0.44}
\newcommand{\gpMADnp}{0.48}

\newcommand{\percentgpMADnp}{2.62}

\newcommand{\telavivRMS}{4.58}
\newcommand{\telavivMAD}{0.37}

\newcommand{\percenttelavivMAD}{1.75}

\title[GP rotation periods]{Inferring probabilistic stellar rotation periods
using Gaussian processes}

\author[R.~Angus \emph{et al.}]{
    Ruth~Angus,$^1$\thanks{Contact e-mail:
\href{mailto:ruthangus@astro.columbia.edu}{ruthangus@astro.columbia.edu}}
    Timothy Morton$^2$,
    Suzanne Aigrain$^3$,
    Daniel Foreman-Mackey$^4$
    \newauthor
    \& Vinesh Rajpaul$^3$
    \\
    $^1$Simons Fellow, Department of Astronomy, Columbia University, NY, NY \\
    $^2$Department of Astrophysical Sciences, Princeton University,
    Princeton, NJ \\
    $^3$Subdepartment of Astrophysics, University of Oxford, UK \\
    $^4$Sagan Fellow, Department of Astronomy, University of Washington,
    Seattle, WA}

\date{Last updated 2017 March 13; in original form 2017 March 13}
\pubyear{2017}
\begin{document}
\label{firstpage}
\pagerange{\pageref{firstpage}--\pageref{lastpage}}
\maketitle

\begin{abstract}
Variability in the light curves of spotted, rotating stars is often
    non-sinusoidal and quasi-periodic --- spots move on the stellar surface
    and have finite lifetimes, causing stellar flux variations to slowly shift
    in phase.
A strictly periodic sinusoid therefore cannot accurately model a rotationally
    modulated stellar light curve.
Physical models of stellar surfaces have many drawbacks preventing effective
    inference, such as highly degenerate or high-dimensional parameter spaces.
In this work, we test an appropriate {\it effective} model: a Gaussian
    Process with a quasi-periodic covariance kernel function.
This highly flexible model allows sampling of the posterior probability
    density function of the periodic parameter, marginalising over the
    other kernel hyperparameters using a Markov Chain Monte Carlo approach.
To test the effectiveness of this method, we infer rotation periods from
    \naigrain\ simulated stellar light curves, demonstrating that the Gaussian
    process method produces periods that are more accurate than both
    a sine-fitting periodogram and an autocorrelation function method.
We also demonstrate that it works well on real data, by inferring
    rotation periods for \nkoimcq\ \Kepler\ stars with previously measured
    periods.
    \response{We provide a table of rotation periods for these \nkoi\ \Kepler\ objects of
    interest and their posterior probability density function samples.}
Because this method delivers posterior probability density functions, it will
    enable hierarchical studies involving stellar rotation, particularly those
    involving population modelling, such as inferring stellar ages,
    obliquities in exoplanet systems, or characterising star-planet
    interactions.
    The code used to implement this method is available online.

\end{abstract}


\section{Introduction}
\label{sec:intro}

The disk-integrated flux of a spotted, rotating star often varies in a
non-sinusoidal and Quasi-Periodic (QP) manner, due to active regions on its
surface which rotate in and out of view.
Complicated surface spot patterns produce non-sinusoidal variations, and the
finite lifetimes of these active regions and differential rotation on the
stellar surface produce quasi-periodicity \citep{Dumusque2011}.
A strictly periodic sinusoid is therefore not necessarily a good model choice
for these time-series.
A physically realistic model of the stellar surface
would perfectly capture the complexity of shapes
within stellar light curves as well as the quasi-periodic nature, allowing for
extremely precise probabilistic period recovery when conditioned on the data.
However, such physical models require many free parameters in order to
accurately represent a stellar surface, and some of these parameters are
extremely degenerate \citep[\eg][]{Russell1906, Jeffers2009, Kipping2012}.
In addition to global stellar parameters such as inclination and rotation
period, each spot or active region should have (at minimum) a longitude,
latitude, size, temperature and lifetime.
Considering that stars may have hundreds of spots, the number of free
parameters in such a model quickly becomes unwieldy, especially to explore its
posterior Probability Density Function (PDF).
Simplified spot models, such as the one described in \citet{Lanza2014} where
only two spots are modelled, have produced successful results; however, such
relatively inflexible models sacrifice precision.

Standard non-inference based methods to measure rotation periods include
detecting peaks in a Lomb-Scargle \citep{Lomb1976, Scargle1982} (LS)
periodogram \citep[\eg][]{Reinhold2013}, Auto-Correlation Functions (ACFs),
\citep[\eg][]{Mcquillan13b} and wavelet transforms \citep[\eg][]{Garcia2014}.
The precisions of the LS periodogram and wavelet methods are limited by the
suitability of the model choice: a sinusoid for the LS periodogram, and a
choice of mother wavelet, assumed to describe the data over a range of
transpositions and scales \citep[see, \eg][]{Carter2010}, for the wavelet
method.
In contrast, since it does not rely on a generative model, the ACF method is
much better suited to signals that are non-sinusoidal.
In fact, as long as the signal is approximately periodic the ACF will
display a peak at the rotation period, no matter its shape.
A drawback of the ACF method however, is that it requires data to be
evenly-spaced.
Most ground-based measurements, such as the future Large Synoptic Survey
Telescope (\LSST), do not nearly satisfy this criterion, and even
light curves from the \Kepler\ space telescope, which has provided the richest
stellar rotation data to date, can only be approximated as uniformly sampled.
An ACF is also an operation performed on the data rather than a generative
model of the data, and so is not inherently probabilistic.
This means that the effects of the observational uncertainties cannot be
formally propagated to constraints on the rotation period.

In this work, we test an {\it effective} model for rotationally modulated
stellar light curves which captures the salient behaviour but is not
physically motivated --- although some parameters may indeed be {\it
interpreted} as physical ones.
An ideal effective model should have a small number of non-degenerate
parameters and be flexible enough to perfectly capture non-sinusoidal and QP
behaviour.
A Gaussian process (GP) model fulfills these requirements. We thus use a GP
as the generative model at the core of a method to probabilistically
infer accurate and precise stellar rotation periods.
This enables us to estimate the posterior PDF of the rotation period, thereby
producing a justified estimate of its uncertainty.
\response{Using a quasi-periodic GP model to infer a rotation period is not a new
idea.
For example, we used this method in \citet{Vanderburg2015} to measure the
rotation period of an exoplanet host.
Previously to that, \citet{Haywood2014} used the quasi-periodic GP model
to disentangle the correlated noise produced by stellar activity and rotation
from the radial velocity signature of an exoplanet.
In addition, \citet{Littlefair2017} use a quasi-periodic GP to establish the
quasi-periodicity of variability in the light curves of brown dwarfs.
What we present here is a test of the GP method and a comparison with
alternative rotation period measurement methods.
We also provide code allowing others to easily apply this technique
themselves.}

GPs are commonly used in the machine learning community and increasingly
in other scientific fields such as biology, geophysics and cosmology.
More recently, GPs have been used in the stellar and exoplanet fields within
astronomy, to capture stellar variability or instrumental systematics
\citep[see \eg][]{Gibson2012, Haywood2014, Dawson2014, Barclay2015,
Haywood2015, Evans2015, Rajpaul2015, Czekala2015, Vanderburg2015, Rajpaul2016,
Aigrain2016, Littlefair2017}.
They are useful in regression problems involving any stochastic process,
specifically when the probability distribution for the process is a
multi-variate Gaussian.
If the probability of obtaining a dataset is a Gaussian in $N$ dimensions,
where $N$ is the number of data points, a GP can describe that dataset.
An in-depth introduction to Gaussian processes is provided in
\citet{Rasmussen2005}.

\begin{figure}
\begin{center}
\includegraphics[width=\columnwidth, clip=true]{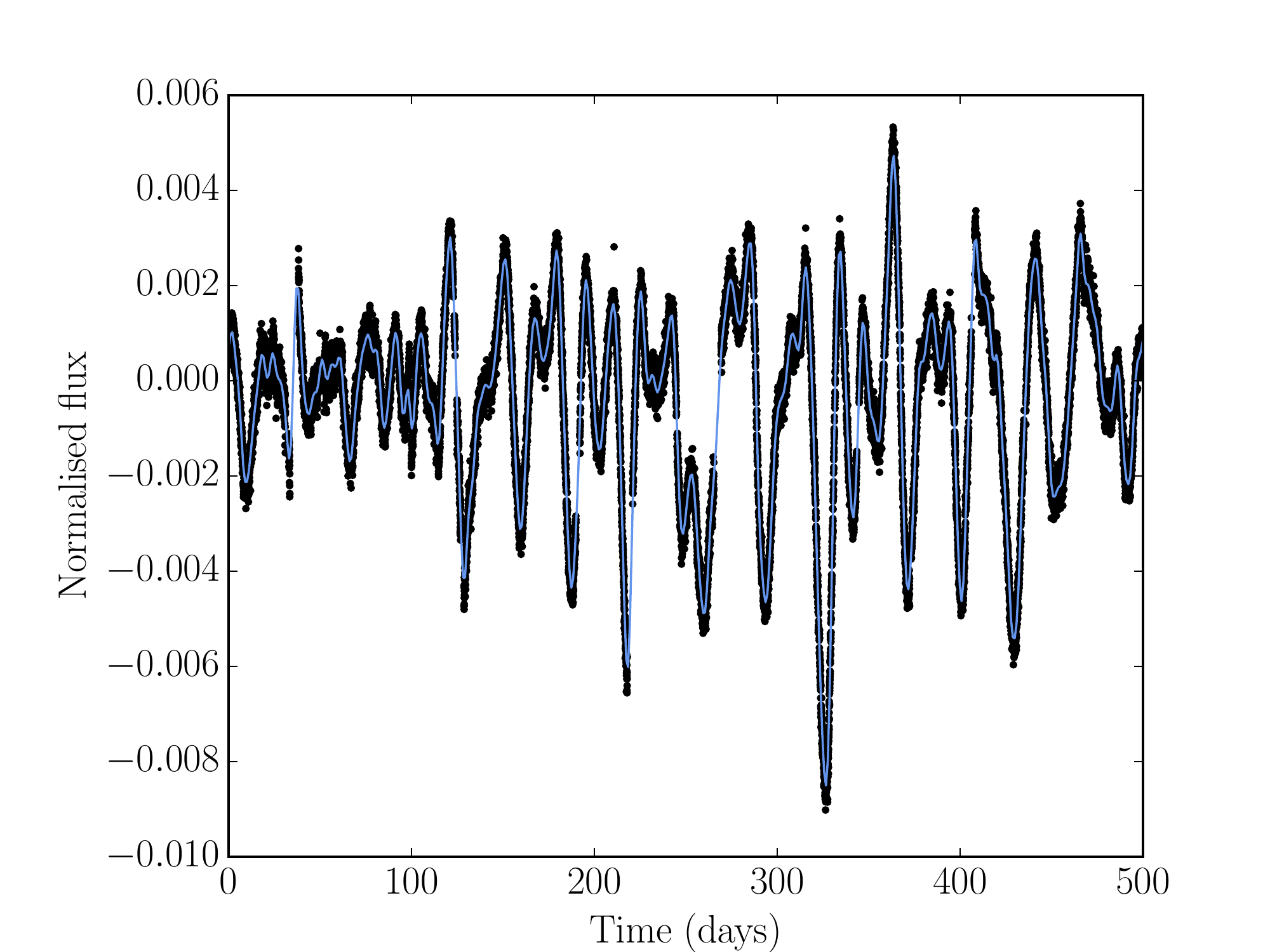}
\caption[A light curve with a GP model.]
{Light curve of KIC \kepexample, an active star with a rotation period of
$\sim$ \kepexampleperiod\ days.
The blue line shows a fit to the data using a Gaussian process model with a QP
covariance kernel function.}
\label{fig:GP_example}
\end{center}
\end{figure}

GP models parameterise the covariance between data points by means of a
kernel function.
As a qualitative demonstration, we present the time-series in figure
\ref{fig:GP_example}: the \kepler\ light curve of KIC \kepexample.
This is a relatively active star that rotates once every $\sim$
\kepexampleperiod\ days, with stochastic variability typical of \kepler\ FGK
stars.
Clearly, data points in this light curve are correlated.
Points close together in time are tightly correlated, and points more
widely separated are loosely correlated.
A GP models this variation in correlation as a function of the separation
between data points; that is, it models the {\it covariance structure} rather
than the data directly.
This lends GPs their flexibility---they can model any time
series with a similar covariance structure.
In addition, a very simple function can usually capture the covariance
structure of a light curve, whereas modelling the time series itself
might require much more complexity.
Figure \ref{fig:GP_example} demonstrates how a GP model fits the light curve of
KIC \kepexample.

A range of covariance models, or kernel functions, can describe stellar
variability.
For example, the most commonly used kernel function, the `Squared
Exponential' (SE), defined as follows, could adequately fit the
KIC \kepexample\ light curve:
\begin{equation}
\label{eq:SE}
k_{i,j} = A \exp \left(-\frac{(x_i - x_j)^2}{2l^2} \right).
\end{equation}
Here $A>0$ is the amplitude of covariance, $l$ is the length scale of
exponential decay, and $x_i-x_j$ is the separation between data points.
The SE kernel function has the advantage of being very simple, with
just two parameters, $A$ and $l$.
If $l$ is large, two data points far apart in $x$ will be tightly correlated,
and if small they will be loosely correlated.
Another property of the SE kernel function is that it produces functions that
are infinitely differentiable, making it possible to model a data set
and its derivatives simultaneously.
However, The SE kernel function does not well describe the covariance
in stellar light curves, nor is it {\it useful} for the problem of
rotation period inference because it does not capture periodic behaviour.
Inferring rotation periods thus requires a periodic kernel
function.
For this reason, we use the `Quasi-Periodic' kernel.
\citet{Rasmussen2005} model QP variability in CO$_2$ concentration on the
summit of the Mauna Loa volcano in Hawaii \citep[data from][]{Keeling2004}
using a kernel which is the product of a periodic and a SE kernel: the QP
kernel.
This kernel is defined as
\begin{equation}
\label{eq:QP}
k_{i,j} = A \exp \left[-\frac{(x_i - x_j)^2}{2l^2} -
    \Gamma^2 \sin^2\left(\frac{\pi(x_i - x_j)}{P}\right) \right] + \sigma^2
    \delta_{ij}.
\end{equation}
It is the product of the SE kernel function, which describes the overall
covariance decay, and an exponentiated, squared, sinusoidal kernel function
that describes the periodic covariance structure.
$P$ can be interpreted as the rotation period of the star, and $\Gamma$
controls the amplitude of the $\sin^2$ term.
If $\Gamma$ is very large, only points almost exactly one period away are
tightly correlated and points that are slightly more or less than one period
away are very loosely correlated.
If $\Gamma$ is small, points separated by one period are tightly
correlated, and points separated by slightly more or less are still highly
correlated, although less so.
In other words, large values of $\Gamma$ lead to periodic variations with
increasingly complex harmonic content.
This kernel function allows two data points that are separated in time by one
rotation period to be tightly correlated, while also allowing
points separated by half a period to be weakly correlated.
The additional parameter $\sigma$ captures white noise by adding
a term to the diagonal of the covariance matrix.
This can be interpreted to represent underestimation of observational
uncertainties~---~if the uncertainties reported on the data are too small, it
will be non-zero~---~or it can capture any remaining ``jitter,'' or residuals
not captured by the effective GP model.
We use this QP kernel function (Equation \ref{eq:QP}) to produce
the GP model that fits the \Kepler\ light curve in figure
\ref{fig:GP_example}.

There are many ways to construct a QP kernel function, involving a range of
choices for both the periodic and aperiodic components of the model.
The kernel function presented above reproduces the behaviour of stellar light
curves, but other model choices can do so as well.
We do not attempt to test any other models in this paper, noting only that
this kernel function provides an adequate fit to the data.
We leave formal comparisons with other kernel function choices to a future
publication.

To infer a stellar rotation period $P$ from a light curve, we fit this
QP-kernel GP model to the data.  As with any model-fitting exercise, the
likelihood
of the model could be maximised to find the maximum-likelihood value for $P$.
In this study, however, we explore the full posterior PDFs using a Markov
Chain Monte Carlo (MCMC) procedure.  While
this approach comes at a computational cost, such posterior exploration
importantly provides a justified uncertainty estimate.

\response{Performing inference with GPs is computationally expensive.
Is it worth spending the extra computation power to get slightly more
accurate, probabilistic rotation periods for a large number of stars?
A growing number of astronomers are interested in hierarchical probabilistic
modelling, particularly in the stellar and exoplanet communities.
These individuals are mindful of carefully inferring probabilistic parameters
wherever possible.
Probabilistic inference allows one to marginalise over nuisance
parameters while formally taking any uncertainty into account when
measuring the quantity of interest -- the star's rotation period, in
this case.
This becomes especially important as the signal-to-noise ratio
decreases and the rotation period is no longer tightly constrained,
and as the sampling becomes uneven.
Throughout the astronomical literature, hierarchical Bayesian
inference has been demonstrated to be a robust method of population
inference \citep[\eg][]{Hogg2010, Foremanmackey2014, Wolfgang2015, Rogers2015}.
Probabilistic measurements with full, rigorous propagation of
uncertainties are a crucial ingredient for any hierarchical inference
where the target is the population level distribution.
Inferring the ages of stars in the galaxy via gyrochronology (age-rotation
relations) \citep[\eg][]{Skumanich1972, Kawaler1989, Barnes2003, Barnes2007}
is an example of a hierarchical inference problem using stellar rotation.
Hierarchical inference allows you to marginalise over individual rotation
periods and infer age distribution parameters directly from the light curves.
Probabilistic rotation period inference is highly relevant to a large number
of galactic archaeology and exoplanet population studies which require
rotation period (or age) distributions.}

This paper is laid out as follows.
The GP method is described in \textsection \ref{sec:method}.
Its performance is demonstrated and compared with literature methods in
in \textsection \ref{sec:perf_and_comp}.
In \textsection \ref{sec:kepler} we apply the GP method to real \kepler\ data,
and the results are discussed in \textsection \ref{sec:discussion}.

\section{GP Rotation Period Inference}
\label{sec:method}

In order to recover a stellar rotation period from a light curve using a
quasi-periodic Gaussian process (QP-GP), we sample the following posterior
PDF:
\begin{equation}
\label{eq:posterior}
p({\bm \theta}\,|\,y) \propto \mathcal L({\bm \theta}) p({\bm \theta}),
\end{equation}
where $y$ are the light curve flux data, $\bm \theta$ are the hyperparameters
of the kernel described in Equation \ref{eq:QP}, $\mathcal L$ is the
QP-GP likelihood function, and $p({\bm \theta})$ is the prior on the
hyperparameters.  Sampling this posterior presents several challenges:
\begin{itemize}
    \item The likelihood evaluation is computationally expensive;
    \item The GP model is very flexible, sometimes at the expense of
    reliable recovery of the period parameter; and
    \item The posterior may often be multimodal.
\end{itemize}
This Section discusses how we address these challenges through
the implementation details of the likelihood (Section \ref{sec:GP_lhood}), priors
(Section \ref{sec:GP_prior}), and sampling method (Section \ref{sec:sampling}).

\subsection{Likelihood}
\label{sec:GP_lhood}

The GP likelihood is similar to the simple Gaussian likelihood
function where the uncertainties are
Gaussian and uncorrelated. The latter can be written
\begin{equation}
    \ln \mathcal{L} = -\frac{1}{2}\sum_{n=1}^N\left[\frac{(y_n-\mu)^2}{\sigma_n^2}
    + \ln(2\pi\sigma_n^2)\right],
\end{equation}
\label{eq:chi2}
where $y_n$ are the data, $\mu$ is the mean model and $\sigma_n$ are the
Gaussian uncertainties on the data.
The equivalent equation in matrix notation is
\begin{equation}
    \ln \mathcal{L} = -\frac{1}{2}\bf{r}^T\bf{C}^{-1}\bf{r}-\frac{1}{2}\ln|\bf{C}|
    + \frac{N}{2}\log2\pi,
\end{equation}
\label{eq:lhf1}
where $\bf{r}$ is the vector of residuals and $\bf{C}$ is the covariance
matrix,
\begin{eqnarray}
    \mathbf{C} &=& \left (\begin{array}{cccc}
    \sigma^2_1 & \sigma_{2, 1} & \cdots & \sigma_{N, 1} \\
    \sigma_{1, 2} & \sigma^2_2 & \cdots & \sigma_{N, 2} \\
    && \vdots & \\
    \sigma_{1, N} & \sigma_{2, N} & \cdots & \sigma^2_N
\end{array}\right )
\end{eqnarray}
In the case where the uncertainties are uncorrelated, the noise is `white',
(an assumption frequently made by astronomers, sometimes justified)
and the off-diagonal elements of the covariance matrix are zero.
However, in the case where there is evidence for correlated
`noise'\footnote{In our case the `noise' is actually the model!  Incidentally, this approach is the reverse of the regression techniques
usually employed by astronomers.
In most problems in astronomy one tries to infer the parameters that describe
the mean model and, if correlated noise is present, to marginalise over that
noise.
Here, the parameters describing the correlated noise are what we are
interested in and the mean model is simply a straight line at $y=0$.}, as in the
case of \Kepler\ light curves, those off-diagonal elements are non-zero.
With GP regression, a covariance matrix generated by the kernel function
${\bf K}$ replaces ${\bf C}$ in the above equation (for our purposes, the QP
kernel of Equation \ref{eq:QP} generates ${\bf K}$).

Evaluating this likelihood for a large number of points can be computationally
expensive.
For example, evaluating $\mathcal L$ for an entire \Kepler\ lightcurve
($\sim$40,000 points) takes about $\sim$5\,s--- too slow to perform inference
on large numbers of light curves\footnote{All computational times cited in
this section are based on evaluations on a single core of a 2015 Macbook Pro,
3.1 GHz Intel Core i7.}.
The matrix operations necessary to evaluate the GP likelihood scale as
$N\ln(N)^2$, where $N$ is the number of data points in the light curve, using
the fast matrix solver HODLR \citep{Ambikasaran2014},
implemented in the {\tt george} \citep{George} python package.

We accelerate the likelihood calculation using two complementary strategies:
subsampling the data and splitting the light curve into independent sections.
To subsample \Kepler\ data, for example,
we randomly select 1/30th of the points
in the full light curve (an average of $\sim$1.5 points per day).
This decreases the likelihood evaluation time by a factor of about 50, down to
about 100\,ms.
We then split the light curve into equal-sized chunks containing approximately
300 points per section (corresponding to about 200 days), and evaluate the
log-likelihood as the sum of the log-likelihoods of the individual sections
(all using the same parameters ${\bm \theta}$).
This reduces computation time because the section-based likelihood evaluation
scales as $mn\ln(n)^2$, where $n$ is the number of data points per section and
$m$ is the number of sections.
This method further reduces computation time for a typical light curve
(subsampled by a factor of 30) by about a factor of two, to about 50\,ms.

\subsection{Priors}
\label{sec:GP_prior}

\subsubsection{Non-period Hyperparameters}
\label{sec:nonperiod_prior}

The flexibility of this GP model allows for posterior multimodality and
``over-fitting''--like behavior.
For example, if $l$ is small, the non-periodic factor in the covariance
kernel may dominate, allowing for a good fit to the data without
requiring any periodic covariance structure---even if clear periodic
structure is present.
Additionally, for large values of $\Gamma$ the GP model becomes extremely
flexible and can fit the data without varying the period.
Managing this flexibility to reliably retrieve the period parameter requires
imposing informative priors on the non-period GP parameters.
In particular, we find it necessary to avoid large values of
$A$ and $\Gamma$, and small values of $l$; though the exact details of what
works well may differ among datasets.

\subsubsection{Period}
\label{sec:period_prior}

We use an informed period prior, based on the autocorrelation function (ACF)
of the light curve.
The ACF has proven to be very useful for measuring stellar rotation periods
\citep{Mcquillan2012, Mcquillan13b, Mcquillan2014}; however, the
method has several shortcomings,
most notably the inability to deliver uncertainties, but also
the necessity of several heuristic choices,
such as a timescale on which to smooth the ACF,
how to define a peak, whether the first or second peak
gets selected, and what constitutes a secure detection.
While this paper presents a rotation period inference method
that avoids these shortcomings,
it seems prudent to still use information available from the ACF.
We thus use the ACF to define a prior on period,
which can help the posterior sampling converge on the true period.

We do not attempt to decide which single
peak in the ACF best represents the true rotation period,
but rather we identify several \emph{candidate} periods and define
a weighting scheme in order to create a noncommital, though useful,
multimodal prior.  While this procedure does not avoid heuristic choices,
we soften the potential impact of these choices because we are simply
creating a \emph{prior} for probabilistically justified
inference rather attempting to identify a single correct period.

As another innovation beyond what ACF methods in the literature have
presented, we also bandpass filter the light curves (using a 5th order
Butterworth filter, as implemented in \texttt{SciPy}) before calculating the
autocorrelation.  This suppresses power on timescales shorter than a chosen
minimum period $P_{\rm min}$ and longer than a chosen maximum $P_{\rm max}$,
producing a cleaner autocorrelation signal than an unfiltered light curve.

We use the following procedure to construct a prior for rotation period
given a light curve:

\begin{enumerate}
\item{For each value of $P_i$, where $i = \{1, 2, 4, 8, 16, 32, 64, 128\}$\,d,
we apply a bandpass filter to the light curve using $P_{\rm min}=0.1$\,d
and $P_{\rm max} = P_i$.  We then calculate the ACF of the filtered
light curve out to a maximum lag of $2P_i$ and smooth it with a boxcar
filter of width $P_i/10$.}

\item{We identify the time lag corresponding to the
first peak of each of these ACFs, as well as the first peak's
trough-to-peak height, creating a set of candidate periods
$T_i$ and heights $h_i$.}

\item{We assign a quality metric $Q_i$ to each of these candidate
periods, as follows.  First, we model the ACF as a function of lag
time $t$ as a damped oscillator with fixed period $T_i$:
\begin{equation}
y = A e^{-t/\tau} \cos{\frac{2\pi t}{T_i} },
\end{equation}
and find the best-fitting parameters $A_i$ and $\tau_i$ by a non-linear
least squares minimization procedure--though enforcing a maximum
value of $\tau = 20 * P_i/T_i$ to avoid unreasonably
large solutions of $\tau$.
We then define the following heuristic quality metric:
\begin{equation}
\label{eq:quality}
Q_i = \left(\frac{\tau_i}{T_i}\right) \left(\frac{N_i h_i}{R_i}\right),
\end{equation}
where $h_i$ is the height of the ACF peak at $T_i$,
$N_i$ is the length of the lag vector in the ACF (directly proportional
to the maximum allowed period $P_i$),
and $R_i$ is the sum of squared residuals between the
damped oscillator model and the actual ACF data.  The idea behind this
quality metric is to give a candidate period a higher score if

    \begin{enumerate}[(a)]

    \item{it has many regular sinusoidal peaks, such that the decay
        time $\tau_i$ is long compared to the oscillation period $T_i$,}
    \item{the ACF peak height is high, and}
    \item{the damped oscillator model is a good fit (in a $\chi^2$
      sense) to the ACF, with extra bonus for being
      a good fit over more points (larger $N_i$, or longer $P_i$).}

    \end{enumerate}
}

\item{Given this set of candidate periods $T_i$ and quality metrics $Q_i$,
we finally construct a multimodal prior on the $P$ parameter of the GP
model as a weighted mixture of Gaussians:
\begin{multline}
\label{eq:mixture}
p(\ln P) = \\ \frac {\displaystyle \sum_i Q_i \left(0.9\mathcal N(\ln T_i, 0.2) +
                                          0.05\mathcal N(\ln (T_i/2), 0.2) +
                                          0.05\mathcal N(\ln (2 T_i), 0.2) \right)}
                {\sum_i Q_i}.
\end{multline}
That is, in addition to taking the candidate periods themselves as mixture
components, we also mix in twice and half each candidate period at a lower level,
which compensates for the possibility that the first peak in the ACF may actually
represent half or twice the actual rotation period.  The period width of 0.2 in
log space (corresponding to roughly 20\% uncertainty) is again a heuristic choice,
balancing a healthy specificity with the desire to not have the results of
the inference overly determined by the ACF prior.
}
\end{enumerate}

Incidentally, while we use the procedure described here to create a prior on
$P$ which we use while inferring the parameters of the quasi-periodic GP model,
this same procedure may also be used in the service of a
rotation-period estimating procedure all on its own, perhaps being even more
robust and accurate than the traditional ACF method.  We leave exploration of
this possibility to future work.

\subsection{Sampling}
\label{sec:sampling}

To sample the posterior in a way that is sensitive to potential multimodality,
we use the \texttt{emcee3}\footnote{\url{https://github.com/dfm/emcee3}} MCMC
sampler.
\texttt{emcee3} is the successor to the \texttt{emcee} project
\citep{Foreman-Mackey2013} that includes a suite of ensemble MCMC proposals
that can be combined to efficiently sample more distributions than the stretch
move \citep{Goodman2010} in \texttt{emcee}.
For this project, we use a weighted mixture of three proposals.
First, we include a proposal based on the \texttt{kombine}
package\footnote{\url{https://github.com/bfarr/kombine}} (Farr \& Farr, in
prep.) where a kernel density estimate (KDE) of the density represented by the
complementary ensemble is used as the proposal for the other walkers.
The other two proposals are a ``Differential Evolution (DE) MCMC'' proposal
\citep{terBraak2006, Nelson2014} and the ``snooker'' extension of DE
\citep{terBraak2008}.

We initialise 500 walkers with random samples from the prior and use a
weighted mixture of the KDE, DE, and snooker proposals with weights of 0.4,
0.4, and 0.2 respectively.
We run 50 steps of the sampler at a time, checking for convergence
after each iteration, up to a maximum of 50 iterations.
We declare convergence if the total effective chain length is at least
8$\times$ the maximum autocorrelation time.
When convergence is achieved, we discard the first two autocorrelation lengths
in the chain as a burn-in, and randomly choose 5000 samples as representative
of the posterior.
This fitting process takes several hours for a typical simulated light curve,
though in some cases it can take 12 hours or longer to converge.

\section{Performance and Comparison to Literature: Simulated Data}
\label{sec:perf_and_comp}

In order to test this new rotation period recovery method, we apply it to a
set of simulated light curves and compare to the performance of established
literature methods.
Section \ref{sec:simulations} describes the simulated data we use;
Section \ref{sec:performance} demonstrates the performance of the
QP-GP method; and Section \ref{sec:comparison} compares it to the performance
of the Lomb-Scargle periodogram and autocorrelation function methods.

\subsection{Simulated light curves}
\label{sec:simulations}

We take our test data set from the \citet{Aigrain2015} `hare and hounds'
rotation period recovery experiment.
These light curves result from placing dark, circular spots with slowly
evolving size on the surface of bright, rotating spheres, ignoring
limb-darkening effects.
\citet{Aigrain2015} simulated one thousand such light curves to test the
ability of participating teams to recover both stellar rotation periods
and rotational shear (the amplitude of surface differential rotation).
However, in this work, in order to focus on demonstrating reliable period recovery,
we select only the \naigrain\ light curves without differential rotation,
as differential rotation may produce additional scatter in the
measured rotation periods.

\begin{figure}
\begin{center}
\includegraphics[width=\columnwidth, clip=true]{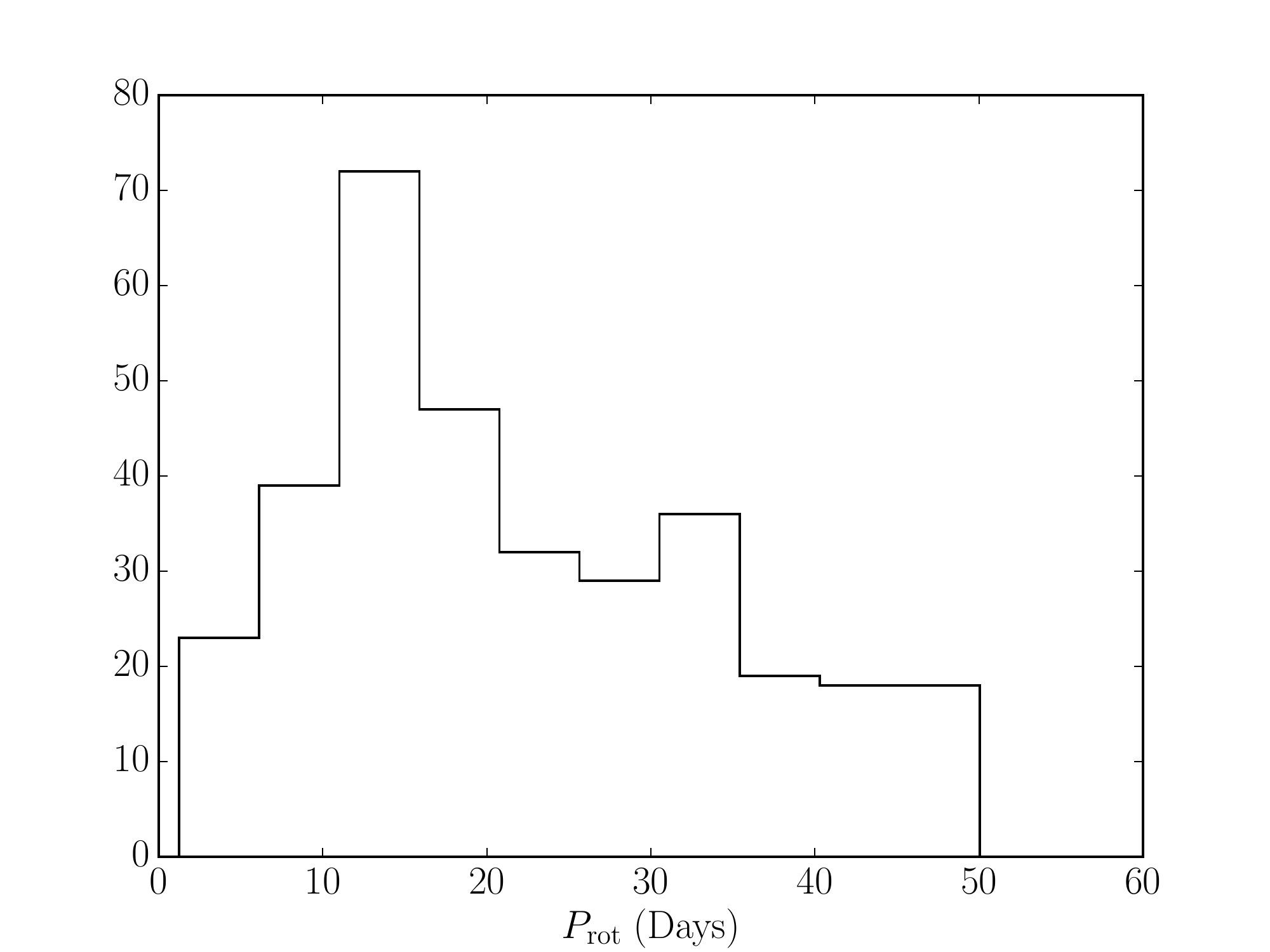}
\caption{A histogram of the rotation periods used to generate the \naigrain\
simulated light curves in \citet{Aigrain2015}.}
\label{fig:period_hist}
\end{center}
\end{figure}

\begin{figure}
\begin{center}
\includegraphics[width=\columnwidth, clip=true]{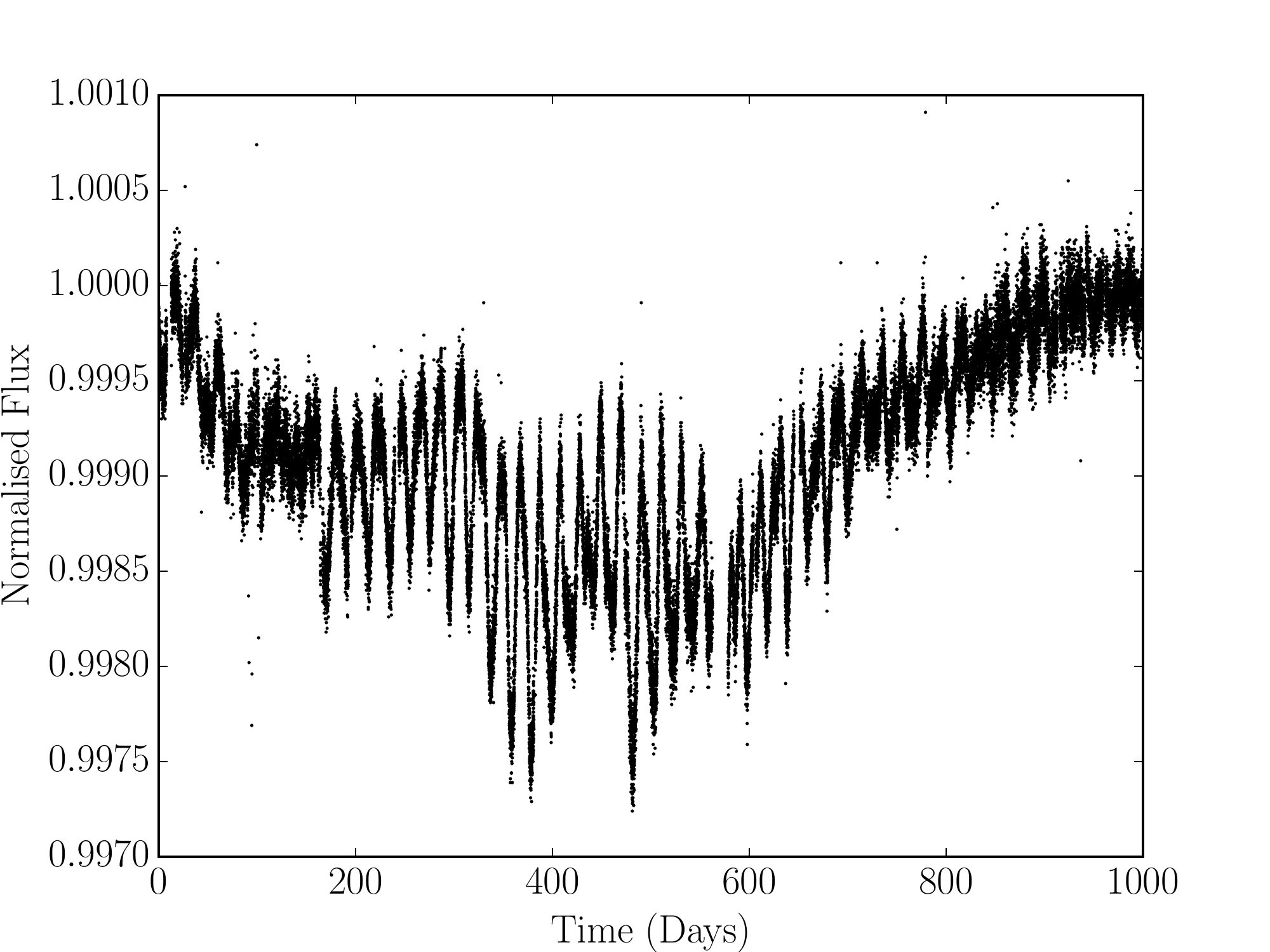}
\caption[A simulated light curve.]
{An example simulated light curve. This `star' has a rotation period of
\aigrainexampleperiod\ days.}
\label{fig:demo_lc}
\end{center}
\end{figure}

Each of these light curve simulations uses a real \Kepler\ long-cadence time array:
one data point every thirty minutes over a four year duration.
90\% of the rotation periods of the simulations come from a
log-uniform distribution between 10 and 50 days, and 10\% from a log-uniform
distribution between 1 and 10 days.
Figure \ref{fig:period_hist} shows the distribution of solid-body rotation periods.
The simulations also have a range of stellar inclination
angles, activity levels, spot lifetimes and more
(see Table \ref{tab:simulation_parameters}).
In order to preserve \kepler\ noise properties,
\citet{Aigrain2015} add real \kepler\ light curves with no obvious astrophysical
variability to the theoretical rotationally modulated light curves.
Figure \ref{fig:demo_lc} shows an example of a simulated light curve with a
period of \aigrainexampleperiod days.

\begin{table*}
\begin{center}
\caption{Ranges and distributions of parameters used to simulate light curves
in \citet{Aigrain2015}}
\begin{tabular}{lcc}
\hline\hline
    Parameter & Range & Distribution \\
    \hline
    Rotation period, $P_{rot}$ & 10 - 50 days (90\%) & log uniform \\
    & 1 - 10 days (10\%) & log uniform \\
    Activity cycle length & 1 - 10 years & log uniform \\
    Inclination & 0 - 90$^\circ$ & Uniform in $\sin^2i$ \\
    Decay timescale & (1 - 10) $\times P_{rot}$ & log uniform \\
\hline
\end{tabular}
\end{center}
\end{table*}
\label{tab:simulation_parameters}

\subsection{Method Performance}
\label{sec:performance}

We apply the QP-GP inference method described in Section \ref{sec:method}
to each of these 333 simulated light curves.  As discussed in Section
\ref{sec:nonperiod_prior}, reliable inference
requires defining a useful set of priors on the non-period
hyperparameters.
For this simulation dataset, we determined these by first running
the method using very broad priors on all the non-period parameters
(log-flat between -20 and 20)
and then inspecting the distribution of their posteriors for those cases
that successfully recovered the true period.
We also experimented with constraining the allowed ranges of the parameters
after discovering that some regions of parameter space (such as large values
of $A$ and $\Gamma$ and small values of $l$) tended to allow fits that ignored
the desired periodicity.
We list the final priors and bounds this process led us to adopt
in Table \ref{tab:priors}.
For the period prior, we tried two different methods: an uninformed (log-flat)
prior between 0.5\,d and 100\,d, and an ACF-informed prior
(Section \ref{sec:period_prior}).

\begin{figure*}
\begin{center}
\includegraphics[width=6in, clip=true]{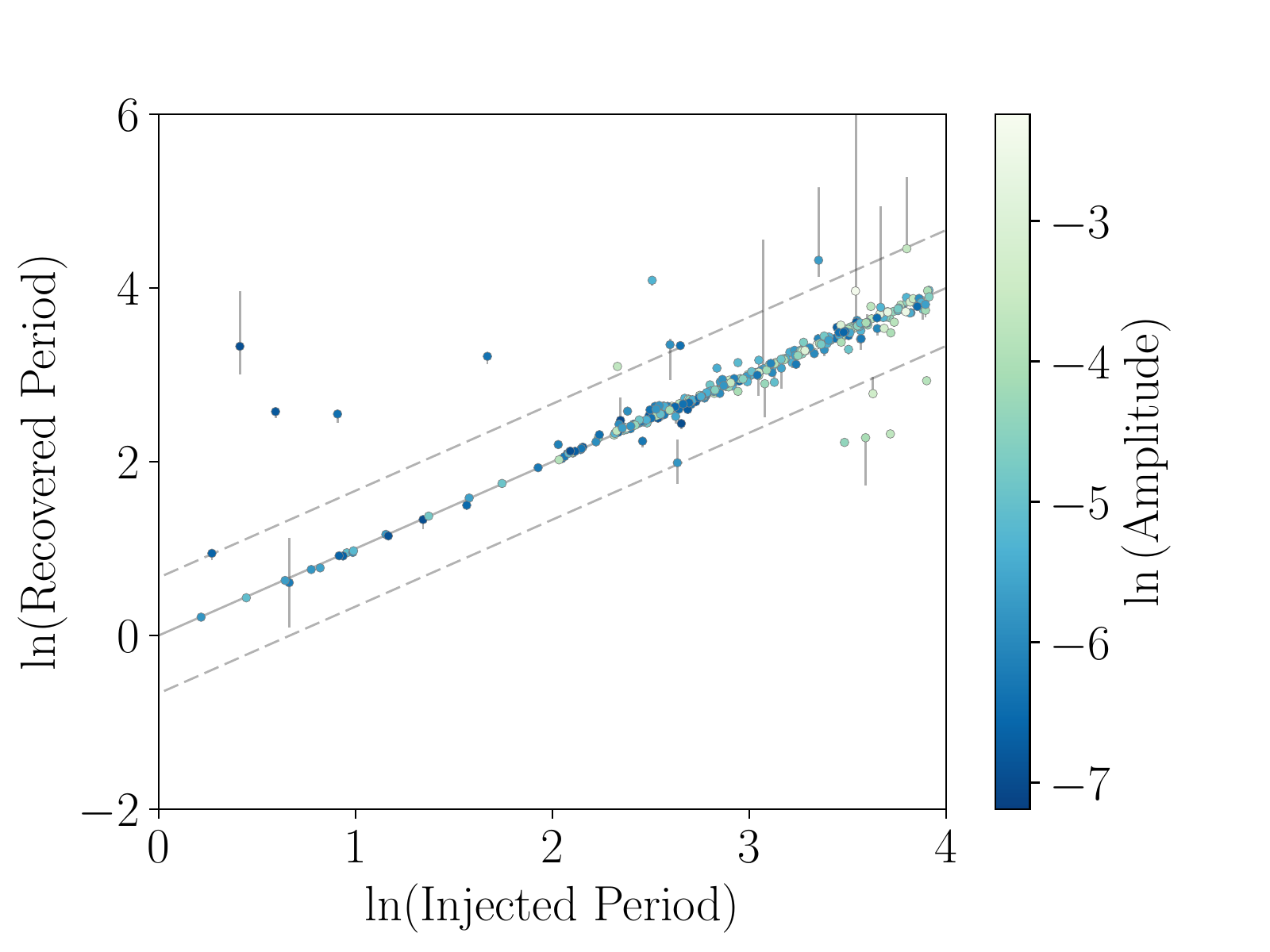}
\caption{The `true' rotation periods used to generate \naigrain\
simulated light curves vs the rotation periods measured using the GP
technique with an informed, ACF-based prior.
    Points are coloured by the peak-to-peak amplitude of the light curve, as
    defined in \citet{Aigrain2015}.
Since the posterior PDFs of rotation periods are often non-Gaussian,
    the points plotted here are maximum {\it a-posteriori} results.
The uncertainties are the 16th and 84th percentiles.
In many cases, the uncertainties are under-estimated.
The ACF-informed prior on rotation period used to generate these results is
    described in \textsection \ref{sec:GP_prior}.
    }
\label{fig:compare_mcmc_acfprior}
\end{center}
\end{figure*}

\begin{figure*}
\begin{center}
\includegraphics[width=6in, clip=true]{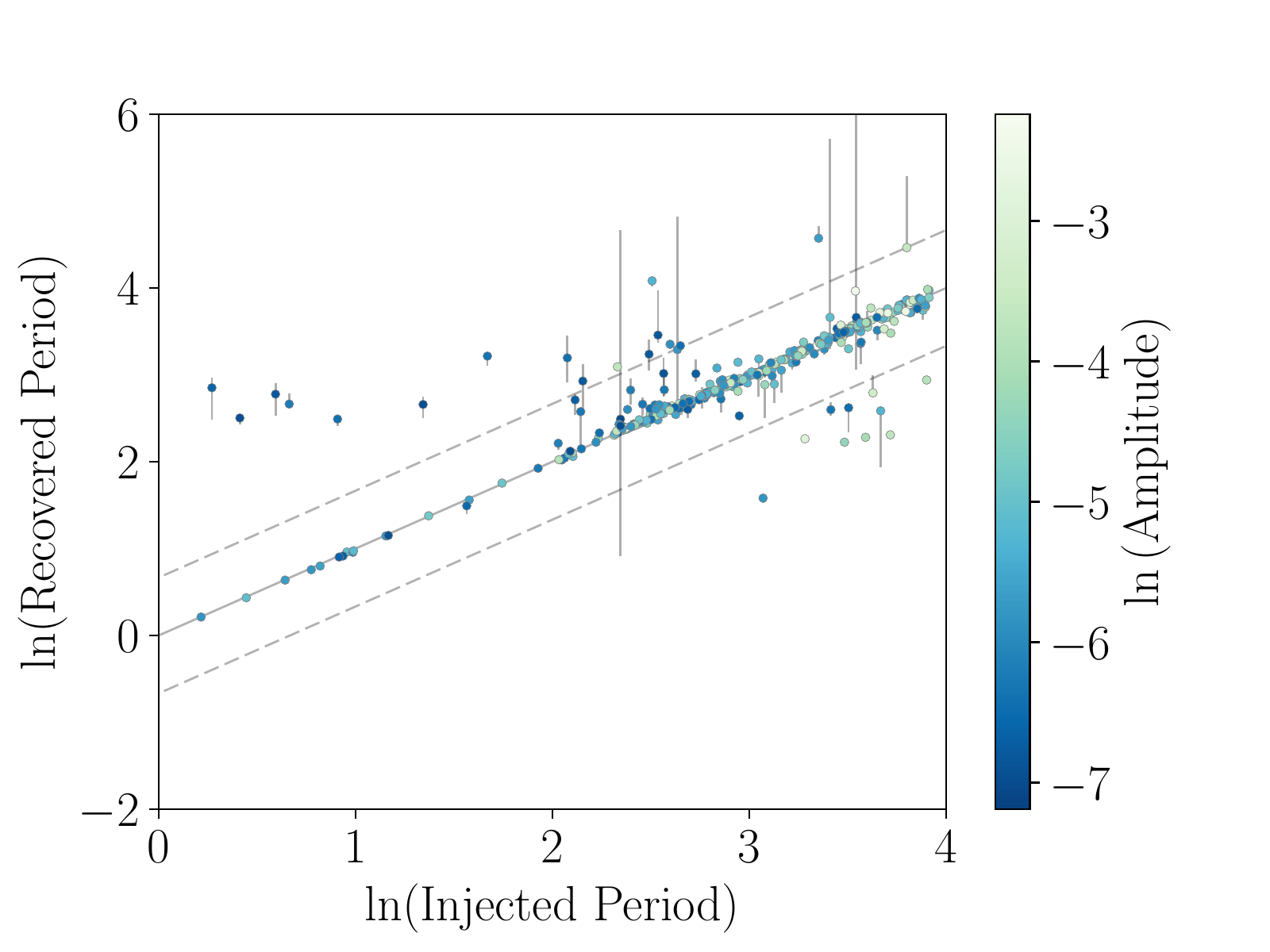}
\caption{The `true' rotation periods used to generate \naigrain\
simulated light curves vs the rotation periods measured using the GP
technique with an uninformative prior.
    Points are coloured by the peak-to-peak amplitude of the light curve, as
    defined in \citet{Aigrain2015}.
Since the posterior PDFs of rotation periods are often non-Gaussian,
    the points plotted here are maximum {\it a-posteriori} results.
The uncertainties are the 16th and 84th percentiles.
In many cases, the uncertainties are under-estimated.
An uninformative prior, flat in the natural log of the rotation period between
    0.5 and 100 days was used to generate these results.
    }
\label{fig:compare_mcmc_noprior}
\end{center}
\end{figure*}

\begin{figure*}
\begin{center}
\includegraphics[width=6in, clip=true]{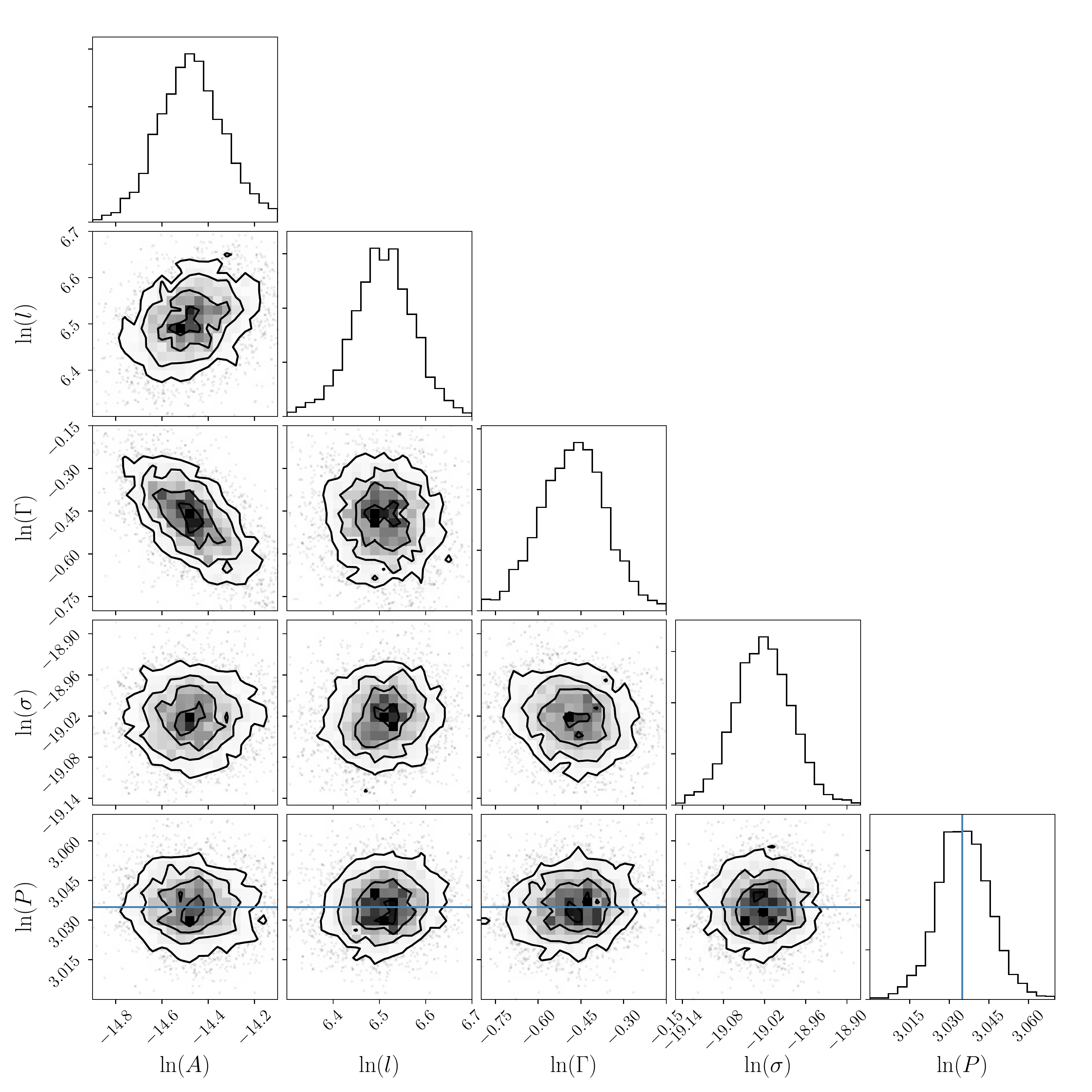}
\caption{Marginal posterior PDFs of the QP GP model parameters, fit to the
    simulated light curve in figure \ref{fig:demo_lc}.
    The blue line in the period panel shows the injected period.
    This figure was made using \texttt{corner.py} \citep{Corner}.}
\label{fig:gp_posteriors}
\end{center}
\end{figure*}

Figures \ref{fig:compare_mcmc_acfprior} and \ref{fig:compare_mcmc_noprior}
summarise our results compared to the injected `true' stellar rotation
periods, for the uninformed and ACF-informed priors on period, respectively.
The periods inferred using the GP method show good agreement with the true
periods.
To assess the performance of the QP-GP and other period recovery methods, we
compute Median Absolute Deviations (MADs) of the results, relative to the
input periods.
We also compute the Median Relative Deviation (MRD), as a percentage and the
root-mean-squared (RMS).
These metrics are presented for the three different methods tested in this
paper in table \ref{tab:MADs}.
The informative and uninformative prior versions of the GP method have MRDs of
\percentgpMAD \% and \percentgpMADnp \% respectively.
The marginal posterior distributions of the QP kernel hyperparameters, for the
example simulated light curve in figure \ref{fig:demo_lc}, are shown in
figure \ref{fig:gp_posteriors}.

\begin{table*}
\begin{center}
\caption{Priors and bounds on the natural logarithms of the GP model
    parameters.}
\begin{tabular}{lcc}
Parameter & Prior & Bounds\\
    \hline
    $\ln A$ & $\mathcal N(-13, 5.7)$ & (-20, 0) \\
    $\ln l$ & $\mathcal N(7.2, 1.2)$ & (2, 20) \\
    $\ln \Gamma$ & $\mathcal N(-2.3, 1.4)$ & (-10, 3) \\
    $\ln \sigma$ & $\mathcal N(-17, 5)$ & (-20, 0) \\
    $\ln P $ & Uniform / ACF-based & ($\ln 0.5, \ln 100$) \\
\end{tabular}
\end{center}
\end{table*}
\label{tab:priors}

\subsection{Comparison with literature methods}
\label{sec:comparison}
\subsubsection{ACF}
\label{sec:acf}


\begin{figure}
\begin{center}
\includegraphics[width=\columnwidth, clip=true]{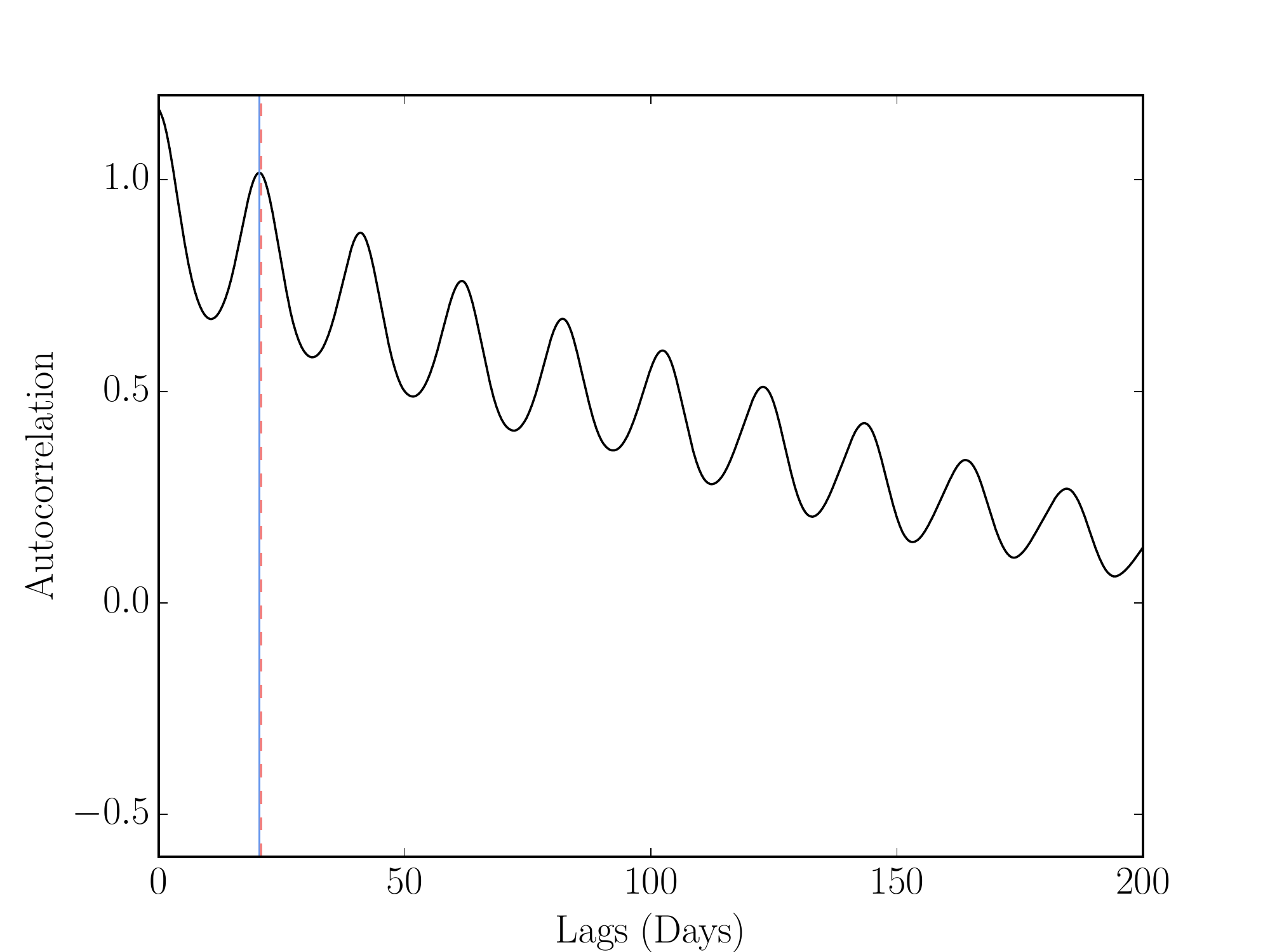}
\caption[ACF of a simulated light curve.]
{An autocorrelation function of the simulated light curve shown in figure
\ref{fig:demo_lc}.
The vertical blue line shows the period measured using the ACF method (20.4
days) and the pink dashed line shows the period that was used to simulate the
light curve (\aigrainexampleperiod\ days).}
\label{fig:demo_acf}
\end{center}
\end{figure}

The ACF method has proven extremely useful for measuring rotation periods.
The catalogue of rotation periods of \Kepler\ stars provided in
\citet{Mcquillan2013} has been widely used by the community and has provided
ground-breaking results for stellar and exoplanetary science.
The method also performed well in the \citet{Aigrain2015}
recovery experiment,
producing a large number of accurate rotation period measurements
(see, e.g., their figure 8).
Another advantage is its fast implementation speed.
However, because the ACF method is non-probabilistic, ACF-estimated rotation
period uncertainties are poorly defined~---~a clear disadvantage.
It also requires evenly-spaced data, therefore is not applicable to
ground-based light curves.
Figure \ref{fig:demo_acf} contains an example ACF of the light curve in
Figure \ref{fig:demo_lc}.

\begin{figure*}
\begin{center}
\includegraphics[width=6in, clip=true]{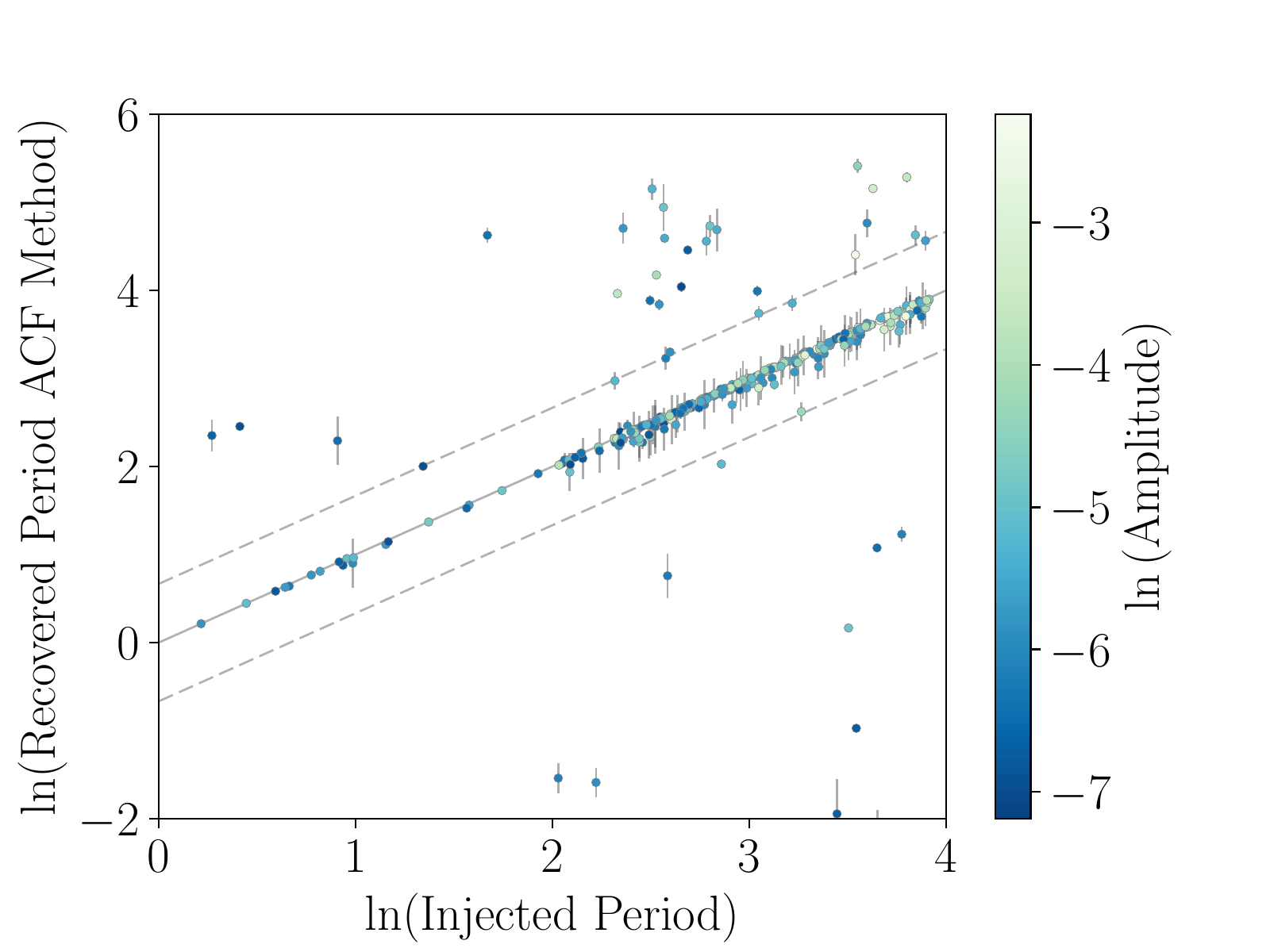}
\caption[ACF results.]
{The `true' rotation periods used to generate \naigrain\ simulated light
curves vs the rotation periods measured using the ACF technique.
    Points are coloured by the peak-to-peak amplitude of the light curve, as
    defined in \citet{Aigrain2015}.
    Several light curves have drastically over- and under-estimated rotation
    periods.
    }
\label{fig:compare_acf}
\end{center}
\end{figure*}

We compare our results to those of the Tel Aviv team in \citet{Aigrain2015}
who use the ACF method described in \citet{Mcquillan2014} and
\citet{Aigrain2015}.
\response{This method involves the calculation of an autocorrelation function, a
measure of the self-similarity of the light curve over a range of lags, which
is then smoothed with a Gaussian kernel.
Periodic variability in the light curve produces a peak in the ACF at the
period of the signal and a decaying sequence of peaks at integer multiples of
the period.
\citet{Mcquillan2014} first pre-processed the light curves by removing very
long-term variations.
They then calculated an ACF for each star and measured the positions of up to
the first four peaks in the ACF.
They fit a straight line to the periods of these peaks as a function of their
integer multiple and adopted the gradient of that line as the rotation
period.
The uncertainty on the ACF period is the uncertainty on the best-fit slope.
}
Figure \ref{fig:compare_acf} shows ACF-measured versus true rotation
periods, with the $2n$ and $\frac{1}{2}n$ harmonic lines as dashed lines.
The injected and recovered rotation periods generally agree well: the MRD of
their periods, relative to the true rotation periods, is \percenttelavivMAD\%
--- slightly smaller than the results produced by the GP method.
However, their results contain a larger number of extreme outliers --- they
predict periods that are significantly different from the true period in more
cases than the GP method.
This is revealed by comparing the RMS of the two sets of results since the RMS
is more sensitive to outliers.
The RMS of the Tel Aviv method results is \telavivRMS\ days and the RMS of GP
method results is \gpRMS\ days.


\subsubsection{LS periodogram}
\label{sec:ls}

The \citet{Aigrain2015} Tel Aviv team were the only group to report periods
for all \naigrain\ simulated stars, and are therefore the only team with which
we compare results.
The other participating groups chose to report only the period measurements in
which they were confident and are likely to have omitted large outliers.
In order therefore to test our results against another commonly used
technique, we implemented a simple LS periodogram method.

\begin{figure*}
\begin{center}
\includegraphics[width=6in, clip=true]{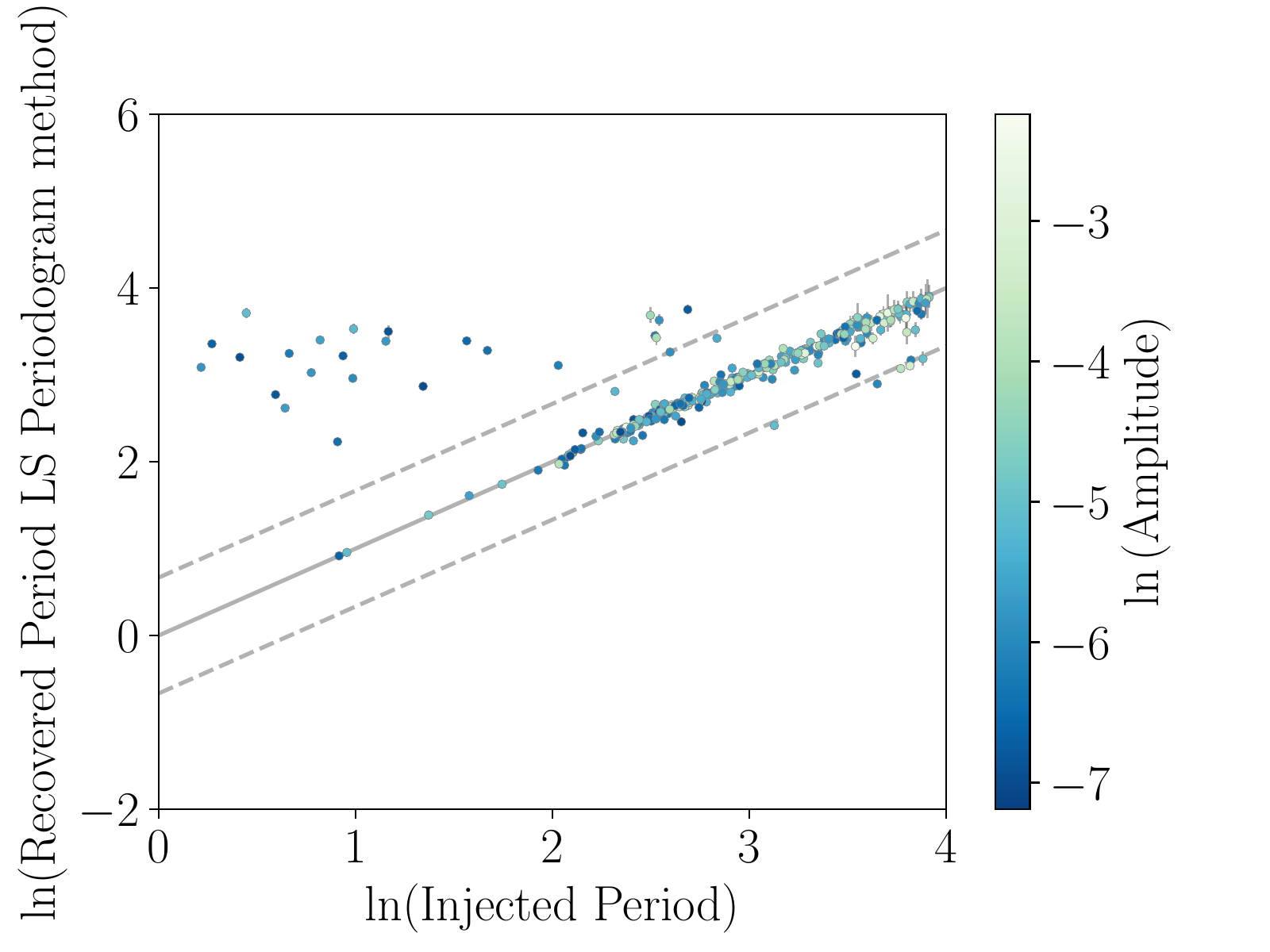}
\caption[LS periodogram results.]
{The `true' rotation periods used to generate \naigrain\ simulated light
curves vs the rotation periods measured using a LS periodogram technique.
    Points are coloured by the peak-to-peak amplitude of the light curve, as
    defined in \citet{Aigrain2015}.
In many cases a large peak at a long period was present in the periodogram,
    producing a significant over-estimate of the period.
    }
\label{fig:pgram_compare}
\end{center}
\end{figure*}

\response{We first applied a high-pass filter to the light curves, removing
long-term trends that could produce large peaks at long periods in the
periodogram.
We used the 3rd order Butterworth filter in {\tt scipy} with a cut-off
of 35 days, attenuating signals with periods greater than this threshold.
This filter was applied in frequency space, rather than period space, and full
attenuation was attained at zero frequency.
In other words, the level of attenuation increased smoothly with period, for
periods above 35 days but no periods were fully suppressed.
Experimentally, we found the 35 day cut off removed the largest number of long
period outliers while preserving shorter period signals.
We also tried 50, 45, 40 and 30 day cut-offs but the 35 day cut-off value
minimised the RMS and MAD of the LS periodogram results.}
For each simulated light curve, we computed a LS periodogram\footnote{LS
periodograms were calculated using the gatspy Python module:
\url{https://github.com/astroML/gatspy/tree/master/gatspy/periodic}.} over a
grid of 10,000 periods, evenly spaced in frequency, between 1 and 100 days.
We adopted the period of the highest peak in the periodogram as the measured
rotation period.
\response{The uncertainties on the rotation periods were calculated using the
following equation for the standard deviation of the frequency
\citep{Horne1986, Kovacs1981}:
\begin{equation}
    \sigma_{\nu} = \frac{3\pi\sigma_N}{2N^{1/2}TA},
\end{equation}
where $A$ is the amplitude of the signal of highest power, $\sigma_N$ is the
variance of the time series, with the signal of highest power removed, $N$
is the number of observations and $T$ is the timespan of the data.
These formal uncertainties are only valid in the case that the noise is white,
the data are evenly sampled and there is only one signal present.
Since there are multiple signals present in these light curves, this formal
uncertainty is an underestimate of the true uncertainty.
}
Figure \ref{fig:pgram_compare} shows the resulting recovered rotation periods
as a function of true period.
\response{The MRD of the periodogram-recovered periods is \percentpgramMAD\%,
slightly worse than the ACF and GP methods (see table \ref{tab:MADs} for
a side-by-side comparison with the other methods).
The filter suppresses signals with periods greater than 35 days but doesn't
eliminate them altogether, thus allowing high amplitude frequencies to be
detectable.
For this reason, some erroneous long periods are recovered by the periodogram
method.
These results still show a marked improvement in comparison to periodograms
calculated from unfiltered data.
Simply adopting the highest peak of a periodogram calculated from the {\it
unfiltered} simulated data resulted in an RMS of \oldpgramRMS\ and relative
MAD of \oldpercentpgramMAD.}

\begin{table*}
\begin{center}
    \caption{Median absolute (MAD), median relative (MRD) deviations and
    root-mean-squared (RMS) for
    the LS periodogram, \citet{Mcquillan2013} ACF and GP period recovery
    methods.}
\begin{tabular}{lcccc}
    Method & MAD & MRD & RMS \\
    \hline
    GP (acf prior) & \gpMAD\ days & \percentgpMAD \% & \gpRMS\ days \\
    GP (uninformative prior) & \gpMADnp\ days & \percentgpMADnp \% &
    \gpRMSnp\ days \\
    ACF & \telavivMAD\ days & \percenttelavivMAD \%
    & \telavivRMS\ days \\
    LS periodogram & \pgramMAD\ days & \percentpgramMAD \% & \pgramRMS\ days\\
\end{tabular}
\end{center}
\end{table*}
\label{tab:MADs}

\section{Real \kepler\ data}
\label{sec:kepler}

In order to test our rotation period inference method on real data,
we apply it to a set of \nkoimcq\ \Kepler\ Object of Interest (KOI)
host stars, that had rotation periods previously measured by
the ACF method by \citet{Mcquillan2013}.
We use the pipeline-corrected flux (\texttt{pdcsap\_flux} column in the
\Kepler\ light curve table), median-normalized and unit-subtracted, and mask
out all known transiting planet candidate signals.
As with the simulated light curves, we randomly subsample each
light curve by a factor of 30 and split it into segments of about 300 points
for the purposes of evaluating the likelihood.
We also follow the same MCMC fitting procedure as with the simulated data,
using the ACF-based prior as before.

Initially, we also use the same priors on the hyperparameters for the KOIs
as for the simulated light curves (Table \ref{tab:priors}).
However, we found that $\ln l$ and $\ln \Gamma$ tended toward slightly
different values than the simulations.
We also found that the allowed hyperparameter range that we used for the
simulations was too large for the KOI population, as maybe $\sim$15\% of the
fits tended toward corners in the hyperparameter space, resulting in poor
period measurements.
As a result, after this initial test, we subsequently adjusted the priors and
re-fit all the KOIs.
The final priors and parameter ranges that we used are in Table
\ref{tab:koipriors}.
The posterior samples of the model parameters for the \Kepler\ objects of
interest are available online:
\url{https://zenodo.org/record/292340\#.WKWpiBIrJE4}.

\begin{table*}
\begin{center}
\caption{Priors and bounds on the natural logarithms of the GP model parameters,
        for \Kepler\ light curves}
\begin{tabular}{lcc}
Parameter & Prior & Bounds\\
    \hline
    $\ln A$ & $\mathcal N(-13, 5.7)$ & (-20, 0) \\
    $\ln l$ & $\mathcal N(5.0, 1.2)$ & (2, 8) \\
    $\ln \Gamma$ & $\mathcal N(1.9, 1.4)$ & (0, 3) \\
    $\ln \sigma$ & $\mathcal N(-17, 5)$ & (-20, 0) \\
    $\ln P $ & ACF-based & ($\ln 0.5, \ln 100$) \\
\end{tabular}
\end{center}
\end{table*}
\label{tab:koipriors}

\begin{figure*}
\begin{center}
\includegraphics[width=6in, clip=true]{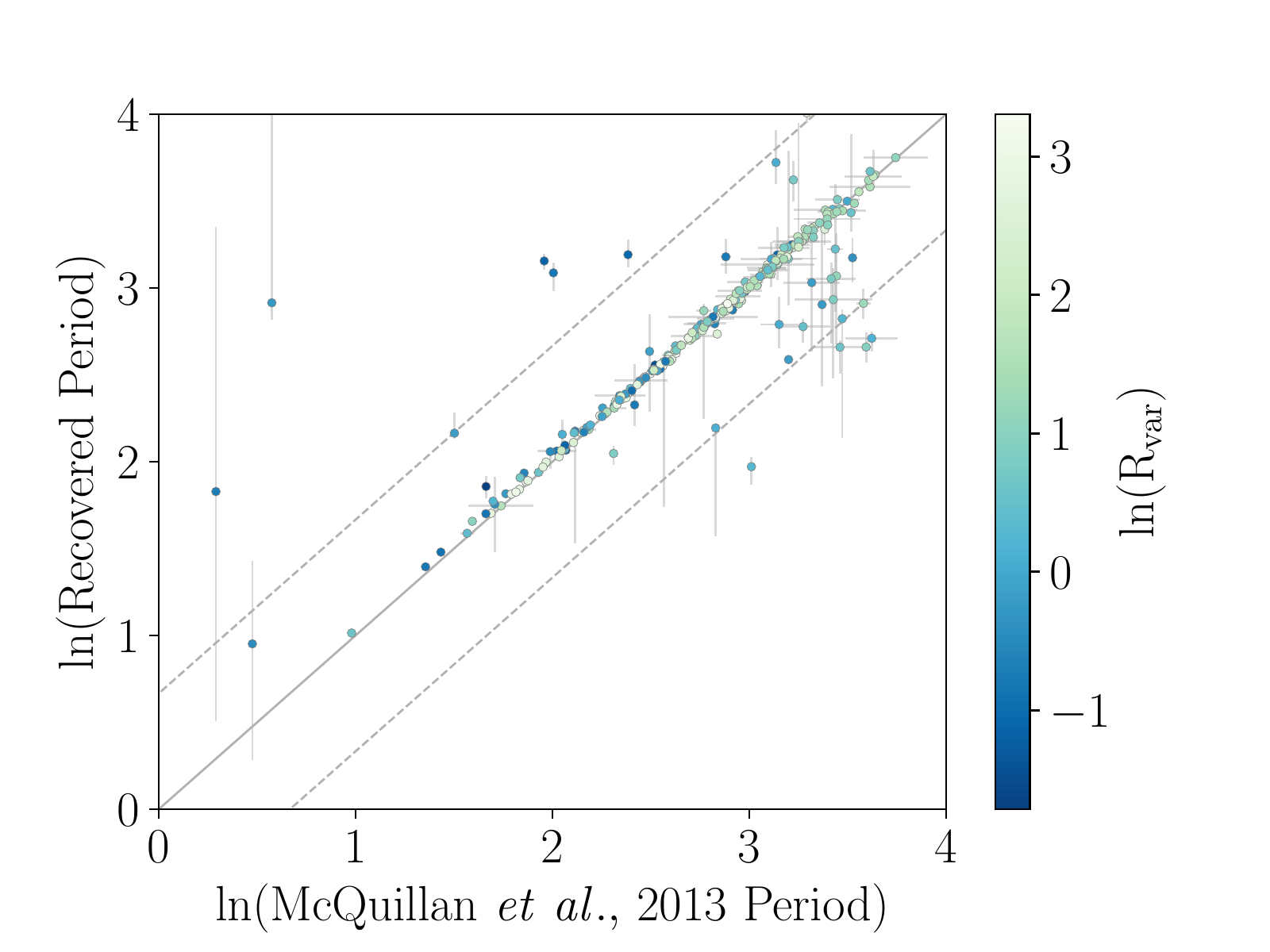}
\caption[Comparison with McQuillan results.]
{A comparison of our GP rotation period measurements to those of
\citet{Mcquillan2013}.
The data points are coloured by the range of variability measured by
    \citet{Mcquillan2013}, defined as the interval between the 5th and 95th
    percentiles of normalised flux per period bin in millimagnitudes.}
\label{fig:mcquillan}
\end{center}
\end{figure*}

Figure \ref{fig:mcquillan} compares the periods inferred with the GP method to
the ACF-based periods from \citet{Mcquillan2013} for the \nkoimcq\ overlapping
KOIs.
This comparison shows generally very good agreement, demonstrating that this
method works not only on simulated data, but also on real data---with the
caveat that for any particular data set, some care is needed regarding the
setting the priors and ranges for the GP hyperparameters.
\response{The new GP rotation periods calculated for \nkoi\ KOIs are provided in an
online table accompanying this paper.
Samples of the rotation period posterior PDFs are available online:
\url{https://doi.org/10.5281/zenodo.804440}}


\section{Discussion}
\label{sec:discussion}

\response{The majority of long period outliers seen in the periodogram comparison
figure (figure \ref{fig:pgram_compare}) are produced when the rotation period
signal is highly non-sinusoidal.
This can be understood intuitively as the LS periodogram fits a series of sine
waves to the data; if the data are not sinusoidal, a sine wave will not
produce a good fit.
It is difficult to assess the reason for the ACF incorrect recoveries because
we do not have the individual ACFs computed by the \citet{Aigrain2015} Tel
Aviv team, however it seems that the ACF method is more sensitive than the LS
periodogram method to noise.
This particular implementation of the LS periodogram method is capable of
detecting periodic signals at lower amplitudes that the ACF.
We do not suggest that the LS periodogram method is better than the ACF method
in general.
It was made clear in \citet{Aigrain2015} that the performance of each of these
rotation period detection algorithms depends sensitively on the
implementation technique.
The exact steps performed in the pre-processing of the light curves and the
post-processing of the periodogram or ACF seem to affect the results strongly.
This is also true for the GP method.
Using a different MCMC sampler, different priors, different subsampling, or a
different kernel would all alter the results.
Clearly, none of these rotation period measurement methods is perfect and they
all require a lot of background engineering and tuning.
This is simply due to inherent difficulty of inferring periods, or any
parameter that produces a highly multimodal probability distribution, from
noisy data.}

The QP-GP inference method we test in this paper produces slightly more
accurate rotation periods than the ACF or periodogram methods.
Additionally, because it explores the posterior PDF with MCMC, the GP method
produces probabilistically justified uncertainties.
Unfortunately, these uncertainties still appear to be underestimated.
Of the rotation periods recovered from the simulated light curves using the
ACF-informed prior, only 25\% of the measured periods lie within 1$\sigma$ of
the true period, 50\% lie within 2$\sigma$, and 66\% within 3$\sigma$.
The largest outlier is 114$\sigma$ away from the true value.
The median uncertainty is underestimated by around a factor of two.
These numbers are similar for the uninformative prior results.
We attribute these underestimated uncertainties to the model choice.
Although more appropriate than a sinusoid, the QP-GP is still only an
 effective model.
A perfect physical model of the star would produce more accurate
uncertainties.

The QP kernel function represents a simple effective model of a stellar
light curve.
It can describe a wide range of quasi-periodic behavior and is
relatively simple, with only a few hyperparameters.
Nevertheless, it is still a somewhat arbitrary choice.
Another valid choice would be a squared cosine function multiplied by a
squared exponential, used by \citet{Brewer2009} to model asteroseismic
pulsations,
\begin{equation}
k_{i,j} = A \exp \left(-\frac{(x_i - x_j)^2}{2l^2}\right)
\cos\left(\frac{\pi(x_i - x_j)}{P}\right).
\end{equation}
\label{eq:cos_kernel}
This function produces a positive semi-definite matrix and has the $P$
parameter of interest, but differs qualitatively from the QP function
by allowing negative covariances.
Is it realistic to allow negative covariances?
In practice, the ACFs of \Kepler\ light curves often go negative.
However, many stars have two active regions on opposite hemispheres that
produce two brightness dips per rotation, and it
may be difficult to model such stars with a kernel that forces
anti-correlation of points $\frac{1}{2}$ a period apart.
It would be worthwhile to test this assumption and this alternative
kernel function (and others) in the future.

Not all \kepler\ light curves show evidence of stellar rotation.
In some cases a star may have few or no active regions, be rotating
pole-on, or be rotating so slowly that the \kepler\ data detrending pipeline
removes any signal.
In other cases there may be another source of variability present in the light
curve, generating a false period detection.
These sources may be physical: \eg\ binary star interactions, intra-pixel
contamination from other astrophysical objects, pulsating variable stars,
asteroseismic oscillations in giants and even stellar activity cycles.
Identifying many of these astrophysical false positives
falls outside the scope of this GP method
(\eg\ applying colour cuts to avoid giant contamination);
however, for some, such as variable stars, they may have distinctive
hyperparameters (\eg\ long coherence timescales) that identify them.
Testing this may be an interesting follow-up study.
As well as astrophysical contamination, instrumental sources may contribute
to contaminating variability, \eg\ temperature variations or pointing shifts
of the \kepler\ spacecraft.
These are unlikely to be periodic and, again, may produce unusual combinations
of hyperparameters.
We also hope to test this in the future.

In addition, we are continuing to develop several other aspects of this GP
method:
\begin{itemize}
    \item{\response{To use an updated method for calculating the GP likelihood and
    drastically improve computational efficiency.
        {\tt celerite}, a new method for rapid computation of the `solve' and
        determinant calculations of covariance matrices necessary for GP
        inference has recently been developed \citep{Foremanmackey2017}.
        {\tt celerite} scales with the number of data points, $\mathcal{O}(N)$
        rather than the $N\log(N)^2$ scaling of the {\tt HODLR} algorithm used
        in this analysis.
        It requires a slightly different functional form for the kernel as the
        increased speed is obtained by using mathematical properties of
        exponential functions.
        Kernels must be written as the sums of exponentials in order to be
        implemented in {\tt celerite}, however a close approximation to the
        kernel function used here is demonstrated to recover rotation periods
        in \citet{Foremanmackey2017}.}}
\item{To investigate more physically motivated priors.
        In this paper we have only explored the physical interpretation of one
        parameter in the kernel function ($P$), which we related to stellar
        rotation periods, but others also warrant physically motivated priors.
        For example, the harmonic complexity parameter, $\Gamma$, defines the
        typical number of inflection points per function period; suitable
        priors may be informed by the typical properties of observed light
        curves.
        For the evolution time-scale, $l$,  it is sensible to enforce $l > P$,
        such that functions that have clear periodicity.
        Intuitively, if $l < P$, functions may evolve significantly over time
        scales shorter than one period, so a function $f(t+P)$ would look
        significantly different to $f(t)$, in which case any claims of
        periodicity become dubious.
        More stringent criteria may be suggested by stellar astrophysical
        considerations, e.g. knowledge of typical active region evolution
        time-scales or persistence lifetimes (e.g. $l\sim2P$).
        Finally, the amplitude hyper-parameter $A$ may be related to active
        region coverage.
        Unfortunately, because these hyper-parameters control the covariance
        structure of functions, their interpretation is not always
        straightforward; interpretation is further complicated by significant
        degeneracies between the hyperparameters.
        Therefore we defer such considerations, along with strategies for
        efficient optimization or marginalization of hyperparameters, to a
        separate paper (currently in prep.).}
\item{To build in a noise model for \kepler\ data.
        Because it is a {\it generative} model of the data, the GP method
        models the rotation period at the same time as systematic noise
        features.  One can then marginalise over the parameters of the noise
        model.
        This approach would be extremely advantageous for \kepler\ data since
        long-term trends are often removed by the \kepler\ detrending
        pipeline.  Marginalising over the noise model at the same time as
        inferring the parameters of interest will insure that the periodic
        signal is preserved.}
\end{itemize}

\section{Conclusion}

We have attempted to recover the rotation periods of \naigrain\ simulated
\kepler-like light curves for solid-body rotators \citep{Aigrain2015} using
three different methods: a Gaussian process method,
an autocorrelation function method, and a Lomb-Scargle periodogram method.
We demonstrate that the GP method produces the most accurate rotation periods
of the three techniques.
Comparing our results with the results of the \citet{Aigrain2015} Tel Aviv
team who implement the ACF method of \citet{Mcquillan2013},
we find their results to be
slightly more precise in a median sense than the GP method results, but with
a larger number of significant outliers.
Additionally, we measure the rotation periods of \nkoimcq\
\kepler\ objects of interest
using the GP method, and find that these results compare well to those
measured previously by \citet{Mcquillan2013}.
The good agreement between these two sets of results demonstrates that the GP
method works well on real \kepler\ data.
Samples of the full set of \nkoi\ KOI rotation period posterior PDFs can be
found at \url{https://doi.org/10.5281/zenodo.804440}.

Unlike the ACF and LS periodogram methods, the GP method provides posterior
PDF samples which can be used to estimate rotation periods uncertainties.
In general it provides more accurate rotation periods than either method and
produces significantly fewer outliers.
The GP method can also be applied to non-uniformly sampled light curves,
unlike the ACF method.

Although an improvement on competing methods, the uncertainties are still
underestimated by the GP method in many cases because it is an approximate
model.
A quasi-periodic Gaussian process is clearly a good model, as demonstrated
here, however it is still only an effective model, not an accurate physical
one.
It captures the posterior PDF of the periodic component of the covariance
matrix of a time series, not the actual rotation period of a physical star.
The marginalised posterior PDF does not therefore exactly reflect the
probability of a rotation period, given the data.
Similarly, the 16th and 84th percentile ranges calculated from the posterior
PDF of the QP-GP model, are not exactly the uncertainty of a physical
rotation period.
Although underestimated uncertainties are a drawback of this method, they are
at least well motivated and well defined which is not the case for
uncertainties calculated from the ACF and LS periodogram methods.

\response{Although GP inference is slow as it requires a `solve' and a matrix
determinant calculation at every step of an MCMC, a new GP solver, {\tt
celerite} \citep{Foremanmackey2017} shows promise for increasing the
computional speed of GP rotation period inference by several orders of
magnitude.
We intend to integrate {\tt celerite} into our code in the near future.
How much computational time is too much is a question that depends on the
available resources of each individual and their scientific priorities.
GP period inference is trivially parallelisable, so if one has access to
thousands of computer cores, one can infer rotation periods of thousands of
stars in a few hours (perhaps just a few minutes with {\tt celerite}).
For single stars GP rotation inference takes a reasonable amount of time and
has already been used on exoplanet hosts \citep[\eg][]{Haywood2014,
Vanderburg2015}.}

The code used in this project is available at
\url{https://github.com/RuthAngus/GProtation/}.

This research was funded by the Simons Foundation and the Leverhulme Trust.
TDM is supported by NASA grant NNX14AE11G, and acknowledges the
hospitality of the Institute for Advanced Study,
where part of this work was completed.
VR thanks Merton College and the National Research Foundation of South Africa
for financial support.
We would like to thank the anonymous referee who's comments significantly
improved this paper.
Some of the data presented in this paper were obtained from the Mikulski
Archive for Space Telescopes (MAST).
STScI is operated by the Association of Universities for Research in
Astronomy, Inc., under NASA contract NAS5-26555.
Support for MAST for non-HST data is provided by the NASA Office of Space
Science via grant NNX09AF08G and by other grants and contracts.
This paper includes data collected by the Kepler mission. Funding for the
\Kepler\ mission is provided by the NASA Science Mission directorate.

\bibliographystyle{plainnat}
\bibliography{GProtation}

\begin{thebibliography}{49}
\providecommand{\natexlab}[1]{#1}
\providecommand{\url}[1]{\texttt{#1}}
\expandafter\ifx\csname urlstyle\endcsname\relax
  \providecommand{\doi}[1]{doi: #1}\else
  \providecommand{\doi}{doi: \begingroup \urlstyle{rm}\Url}\fi

\bibitem[{Aigrain} et~al.(2015){Aigrain}, {Llama}, {Ceillier}, {Chagas},
  {Davenport}, {Garc{\'{\i}}a}, {Hay}, {Lanza}, {McQuillan}, {Mazeh}, {de
  Medeiros}, {Nielsen}, and {Reinhold}]{Aigrain2015}
S.~{Aigrain}, J.~{Llama}, T.~{Ceillier}, M.~L.~d. {Chagas}, J.~R.~A.
  {Davenport}, R.~A. {Garc{\'{\i}}a}, K.~L. {Hay}, A.~F. {Lanza},
  A.~{McQuillan}, T.~{Mazeh}, J.~R. {de Medeiros}, M.~B. {Nielsen}, and
  T.~{Reinhold}.
\newblock {Testing the recovery of stellar rotation signals from Kepler light
  curves using a blind hare-and-hounds exercise}.
\newblock \emph{\mnras}, 450:\penalty0 3211--3226, July 2015.
\newblock \doi{10.1093/mnras/stv853}.

\bibitem[{Aigrain} et~al.(2016){Aigrain}, {Parviainen}, and
  {Pope}]{Aigrain2016}
S.~{Aigrain}, H.~{Parviainen}, and B.~J.~S. {Pope}.
\newblock {K2SC: Flexible systematics correction and detrending of K2 light
  curves using Gaussian Process regression}.
\newblock \emph{\mnras}, April 2016.
\newblock \doi{10.1093/mnras/stw706}.

\bibitem[{Ambikasaran} et~al.(2014){Ambikasaran}, {Foreman-Mackey},
  {Greengard}, {Hogg}, and {O'Neil}]{Ambikasaran2014}
S.~{Ambikasaran}, D.~{Foreman-Mackey}, L.~{Greengard}, D.~W. {Hogg}, and
  M.~{O'Neil}.
\newblock {Fast Direct Methods for Gaussian Processes}.
\newblock \emph{ArXiv e-prints}, March 2014.

\bibitem[{Barclay} et~al.(2015){Barclay}, {Endl}, {Huber}, {Foreman-Mackey},
  {Cochran}, {MacQueen}, {Rowe}, and {Quintana}]{Barclay2015}
T.~{Barclay}, M.~{Endl}, D.~{Huber}, D.~{Foreman-Mackey}, W.~D. {Cochran},
  P.~J. {MacQueen}, J.~F. {Rowe}, and E.~V. {Quintana}.
\newblock {Radial Velocity Observations and Light Curve Noise Modeling Confirm
  that Kepler-91b is a Giant Planet Orbiting a Giant Star}.
\newblock \emph{\apj}, 800:\penalty0 46, February 2015.
\newblock \doi{10.1088/0004-637X/800/1/46}.

\bibitem[{Barnes}(2003)]{Barnes2003}
S.~A. {Barnes}.
\newblock {On the Rotational Evolution of Solar- and Late-Type Stars, Its
  Magnetic Origins, and the Possibility of Stellar Gyrochronology}.
\newblock \emph{\apj}, 586:\penalty0 464--479, March 2003.
\newblock \doi{10.1086/367639}.

\bibitem[{Barnes}(2007)]{Barnes2007}
S.~A. {Barnes}.
\newblock {Ages for Illustrative Field Stars Using Gyrochronology: Viability,
  Limitations, and Errors}.
\newblock \emph{\apj}, 669:\penalty0 1167--1189, November 2007.
\newblock \doi{10.1086/519295}.

\bibitem[{Brewer} and {Stello}(2009)]{Brewer2009}
B.~J. {Brewer} and D.~{Stello}.
\newblock {Gaussian process modelling of asteroseismic data}.
\newblock \emph{\mnras}, 395:\penalty0 2226--2233, June 2009.
\newblock \doi{10.1111/j.1365-2966.2009.14679.x}.

\bibitem[{Carter} and {Winn}(2010)]{Carter2010}
J.~A. {Carter} and J.~N. {Winn}.
\newblock {Parameter Estimation from Time-Series Data with Correlated Errors: A
  Wavelet-Based Method and its Application to Transit Light Curves}.
\newblock Astrophysics Source Code Library, October 2010.

\bibitem[{Czekala} et~al.(2015){Czekala}, {Andrews}, {Mandel}, {Hogg}, and
  {Green}]{Czekala2015}
I.~{Czekala}, S.~M. {Andrews}, K.~S. {Mandel}, D.~W. {Hogg}, and G.~M. {Green}.
\newblock {Constructing a Flexible Likelihood Function for Spectroscopic
  Inference}.
\newblock \emph{\apj}, 812:\penalty0 128, October 2015.
\newblock \doi{10.1088/0004-637X/812/2/128}.

\bibitem[{Dawson} et~al.(2014){Dawson}, {Johnson}, {Fabrycky},
  {Foreman-Mackey}, {Murray-Clay}, {Buchhave}, {Cargile}, {Clubb}, {Fulton},
  {Hebb}, {Howard}, {Huber}, {Shporer}, and {Valenti}]{Dawson2014}
R.~I. {Dawson}, J.~A. {Johnson}, D.~C. {Fabrycky}, D.~{Foreman-Mackey}, R.~A.
  {Murray-Clay}, L.~A. {Buchhave}, P.~A. {Cargile}, K.~I. {Clubb}, B.~J.
  {Fulton}, L.~{Hebb}, A.~W. {Howard}, D.~{Huber}, A.~{Shporer}, and J.~A.
  {Valenti}.
\newblock {Large Eccentricity, Low Mutual Inclination: The Three-dimensional
  Architecture of a Hierarchical System of Giant Planets}.
\newblock \emph{\apj}, 791:\penalty0 89, August 2014.
\newblock \doi{10.1088/0004-637X/791/2/89}.

\bibitem[{Dumusque} et~al.(2011){Dumusque}, {Santos}, {Udry},
  et~al.]{Dumusque2011}
X.~{Dumusque}, N.~C. {Santos}, S.~{Udry}, et~al.
\newblock {Planetary detection limits taking into account stellar noise. II.
  Effect of stellar spot groups on radial-velocities}.
\newblock \emph{\aap}, 527:\penalty0 A82, March 2011.
\newblock \doi{10.1051/0004-6361/201015877}.

\bibitem[{Evans} et~al.(2015){Evans}, {Aigrain}, {Gibson}, {Barstow},
  {Amundsen}, {Tremblin}, and {Mourier}]{Evans2015}
T.~M. {Evans}, S.~{Aigrain}, N.~{Gibson}, J.~K. {Barstow}, D.~S. {Amundsen},
  P.~{Tremblin}, and P.~{Mourier}.
\newblock {A uniform analysis of HD 209458b Spitzer/IRAC light curves with
  Gaussian process models}.
\newblock \emph{\mnras}, 451:\penalty0 680--694, July 2015.
\newblock \doi{10.1093/mnras/stv910}.

\bibitem[{Foreman-Mackey} et~al.(2013){Foreman-Mackey}, {Hogg}, {Lang},
  et~al.]{Foreman-Mackey2013}
D.~{Foreman-Mackey}, D.~W. {Hogg}, D.~{Lang}, et~al.
\newblock {emcee: The MCMC Hammer}.
\newblock \emph{\pasp}, 125:\penalty0 306--312, March 2013.
\newblock \doi{10.1086/670067}.

\bibitem[{Foreman-Mackey} et~al.(2014{\natexlab{a}}){Foreman-Mackey}, {Hogg},
  and {Morton}]{Foremanmackey2014}
D.~{Foreman-Mackey}, D.~W. {Hogg}, and T.~D. {Morton}.
\newblock {Exoplanet Population Inference and the Abundance of Earth Analogs
  from Noisy, Incomplete Catalogs}.
\newblock \emph{\apj}, 795:\penalty0 64, November 2014{\natexlab{a}}.
\newblock \doi{10.1088/0004-637X/795/1/64}.

\bibitem[{Foreman-Mackey} et~al.(2014{\natexlab{b}}){Foreman-Mackey}, {Hoyer},
  {Bernhard}, and {Angus}]{George}
D.~{Foreman-Mackey}, S.~{Hoyer}, J.~{Bernhard}, and R.~{Angus}.
\newblock {Fast Gaussian Processes for regression}.
\newblock October 2014{\natexlab{b}}.
\newblock \doi{10.5281/zenodo.11989}.
\newblock URL \url{http://dx.doi.org/10.5281/zenodo.11989}.

\bibitem[{Foreman-Mackey} et~al.(2017){Foreman-Mackey}, {Agol}, {Angus}, and
  {Ambikasaran}]{Foremanmackey2017}
D.~{Foreman-Mackey}, E.~{Agol}, R.~{Angus}, and S.~{Ambikasaran}.
\newblock {Fast and scalable Gaussian process modeling with applications to
  astronomical time series}.
\newblock \emph{ArXiv e-prints}, March 2017.

\bibitem[Foreman-Mackey(2016)]{Corner}
Daniel Foreman-Mackey.
\newblock corner.py: Scatterplot matrices in python.
\newblock \emph{The Journal of Open Source Software}, 24, 2016.
\newblock \doi{10.21105/joss.00024}.
\newblock URL \url{http://dx.doi.org/10.5281/zenodo.45906}.

\bibitem[{Garc{\'{\i}}a} et~al.(2014){Garc{\'{\i}}a}, {Ceillier}, {Salabert},
  {Mathur}, {van Saders}, {Pinsonneault}, {Ballot}, {Beck}, {Bloemen},
  {Campante}, {Davies}, {do Nascimento}, {Mathis}, {Metcalfe}, {Nielsen},
  {Su{\'a}rez}, {Chaplin}, {Jim{\'e}nez}, and {Karoff}]{Garcia2014}
R.~A. {Garc{\'{\i}}a}, T.~{Ceillier}, D.~{Salabert}, S.~{Mathur}, J.~L. {van
  Saders}, M.~{Pinsonneault}, J.~{Ballot}, P.~G. {Beck}, S.~{Bloemen}, T.~L.
  {Campante}, G.~R. {Davies}, J.-D. {do Nascimento}, Jr., S.~{Mathis}, T.~S.
  {Metcalfe}, M.~B. {Nielsen}, J.~C. {Su{\'a}rez}, W.~J. {Chaplin},
  A.~{Jim{\'e}nez}, and C.~{Karoff}.
\newblock {Rotation and magnetism of Kepler pulsating solar-like stars. Towards
  asteroseismically calibrated age-rotation relations}.
\newblock \emph{\aap}, 572:\penalty0 A34, December 2014.
\newblock \doi{10.1051/0004-6361/201423888}.

\bibitem[{Gibson} et~al.(2012){Gibson}, {Aigrain}, {Roberts}, {Evans},
  {Osborne}, and {Pont}]{Gibson2012}
N.~P. {Gibson}, S.~{Aigrain}, S.~{Roberts}, T.~M. {Evans}, M.~{Osborne}, and
  F.~{Pont}.
\newblock {A Gaussian process framework for modelling instrumental systematics:
  application to transmission spectroscopy}.
\newblock \emph{\mnras}, 419:\penalty0 2683--2694, January 2012.
\newblock \doi{10.1111/j.1365-2966.2011.19915.x}.

\bibitem[{Goodman, J.} and Weare(2010)]{Goodman2010}
{Goodman, J.} and J.~Weare.
\newblock {Ensemble samplers with affine invariance}.
\newblock \emph{Communications in Applied Mathematics and Computational
  Science}, 5:\penalty0 1, 2010.

\bibitem[{Haywood}(2015)]{Haywood2015}
R.~D. {Haywood}.
\newblock \emph{{Hide and Seek: Radial-Velocity Searches for Planets around
  Active Stars}}.
\newblock PhD thesis, University of St Andrews, November 2015.

\bibitem[{Haywood} et~al.(2014){Haywood}, {Collier Cameron}, {Queloz},
  {Barros}, {Deleuil}, {Fares}, {Gillon}, {Lanza}, {Lovis}, {Moutou}, {Pepe},
  {Pollacco}, {Santerne}, {S{\'e}gransan}, and {Unruh}]{Haywood2014}
R.~D. {Haywood}, A.~{Collier Cameron}, D.~{Queloz}, S.~C.~C. {Barros},
  M.~{Deleuil}, R.~{Fares}, M.~{Gillon}, A.~F. {Lanza}, C.~{Lovis},
  C.~{Moutou}, F.~{Pepe}, D.~{Pollacco}, A.~{Santerne}, D.~{S{\'e}gransan}, and
  Y.~C. {Unruh}.
\newblock {Planets and stellar activity: hide and seek in the CoRoT-7 system}.
\newblock \emph{\mnras}, 443:\penalty0 2517--2531, September 2014.
\newblock \doi{10.1093/mnras/stu1320}.

\bibitem[{Hogg} et~al.(2010){Hogg}, {Myers}, and {Bovy}]{Hogg2010}
D.~W. {Hogg}, A.~D. {Myers}, and J.~{Bovy}.
\newblock {Inferring the Eccentricity Distribution}.
\newblock \emph{\apj}, 725:\penalty0 2166--2175, December 2010.
\newblock \doi{10.1088/0004-637X/725/2/2166}.

\bibitem[{Horne} and {Baliunas}(1986)]{Horne1986}
J.~H. {Horne} and S.~L. {Baliunas}.
\newblock {A prescription for period analysis of unevenly sampled time series}.
\newblock \emph{\apj}, 302:\penalty0 757--763, March 1986.
\newblock \doi{10.1086/164037}.

\bibitem[{Jeffers} and {Keller}(2009)]{Jeffers2009}
S.~V. {Jeffers} and C.~U. {Keller}.
\newblock {An analytical model to demonstrate the reliability of reconstructed
  `active longitudes'.}
\newblock In E.~{Stempels}, editor, \emph{15th Cambridge Workshop on Cool
  Stars, Stellar Systems, and the Sun}, volume 1094 of \emph{American Institute
  of Physics Conference Series}, pages 664--667, February 2009.
\newblock \doi{10.1063/1.3099201}.

\bibitem[{Kawaler}(1989)]{Kawaler1989}
S.~D. {Kawaler}.
\newblock {Rotational dating of middle-aged stars}.
\newblock \emph{\apjl}, 343:\penalty0 L65--L68, August 1989.
\newblock \doi{10.1086/185512}.

\bibitem[{Keeling} and {Whorf}(2004)]{Keeling2004}
D.~{Keeling}, C. and P.~{Whorf}, T.
\newblock {Atmospheric CO2 from Continuous Air Samples at Mauna Loa
  Observatory, Hawaii, U.S.A.}
\newblock \emph{Carbon Dioxide Information Analysis Center, Oak Ridge National
  Laboratory}, October 2004.

\bibitem[{Kipping}(2012)]{Kipping2012}
D.~M. {Kipping}.
\newblock {An analytic model for rotational modulations in the photometry of
  spotted stars}.
\newblock \emph{\mnras}, 427:\penalty0 2487--2511, December 2012.
\newblock \doi{10.1111/j.1365-2966.2012.22124.x}.

\bibitem[{Kovacs}(1981)]{Kovacs1981}
G.~{Kovacs}.
\newblock {Frequency shift in Fourier analysis}.
\newblock \emph{\apss}, 78:\penalty0 175--188, August 1981.
\newblock \doi{10.1007/BF00654032}.

\bibitem[{Lanza} et~al.(2014){Lanza}, {Das Chagas}, and {De
  Medeiros}]{Lanza2014}
A.~F. {Lanza}, M.~L. {Das Chagas}, and J.~R. {De Medeiros}.
\newblock {Measuring stellar differential rotation with high-precision
  space-borne photometry}.
\newblock \emph{\aap}, 564:\penalty0 A50, April 2014.
\newblock \doi{10.1051/0004-6361/201323172}.

\bibitem[{Littlefair} et~al.(2017){Littlefair}, {Burningham}, and
  {Helling}]{Littlefair2017}
S.~P. {Littlefair}, B.~{Burningham}, and C.~{Helling}.
\newblock {Robust detection of quasi-periodic variability: a HAWK-I mini survey
  of late-T dwarfs}.
\newblock \emph{\mnras}, 466:\penalty0 4250--4258, April 2017.
\newblock \doi{10.1093/mnras/stw3376}.

\bibitem[{Lomb}(1976)]{Lomb1976}
N.~R. {Lomb}.
\newblock {Least-squares frequency analysis of unequally spaced data}.
\newblock \emph{\apss}, 39:\penalty0 447--462, February 1976.
\newblock \doi{10.1007/BF00648343}.

\bibitem[{McQuillan} et~al.(2012){McQuillan}, {Aigrain}, and
  {Roberts}]{Mcquillan2012}
A.~{McQuillan}, S.~{Aigrain}, and S.~{Roberts}.
\newblock {Statistics of Stellar Variability in Kepler Data with ARC
  Systematics Removal}.
\newblock In E.~{Griffin}, R.~{Hanisch}, and R.~{Seaman}, editors, \emph{IAU
  Symposium}, volume 285 of \emph{IAU Symposium}, pages 364--365, April 2012.
\newblock \doi{10.1017/S1743921312001081}.

\bibitem[{McQuillan} et~al.(2013{\natexlab{a}}){McQuillan}, {Aigrain}, and
  {Mazeh}]{Mcquillan2013}
A.~{McQuillan}, S.~{Aigrain}, and T.~{Mazeh}.
\newblock {Measuring the rotation period distribution of field M dwarfs with
  Kepler}.
\newblock \emph{\mnras}, 432:\penalty0 1203--1216, June 2013{\natexlab{a}}.
\newblock \doi{10.1093/mnras/stt536}.

\bibitem[{McQuillan} et~al.(2013{\natexlab{b}}){McQuillan}, {Mazeh}, and
  {Aigrain}]{Mcquillan13b}
A.~{McQuillan}, T.~{Mazeh}, and S.~{Aigrain}.
\newblock {Stellar Rotation Periods of the Kepler Objects of Interest: A Dearth
  of Close-in Planets around Fast Rotators}.
\newblock \emph{\apjl}, 775:\penalty0 L11, September 2013{\natexlab{b}}.
\newblock \doi{10.1088/2041-8205/775/1/L11}.

\bibitem[{McQuillan} et~al.(2014){McQuillan}, {Mazeh}, and
  {Aigrain}]{Mcquillan2014}
A.~{McQuillan}, T.~{Mazeh}, and S.~{Aigrain}.
\newblock {Rotation Periods of 34,030 Kepler Main-sequence Stars: The Full
  Autocorrelation Sample}.
\newblock \emph{\apjs}, 211:\penalty0 24, April 2014.
\newblock \doi{10.1088/0067-0049/211/2/24}.

\bibitem[{Nelson} et~al.(2014){Nelson}, {Ford}, and {Payne}]{Nelson2014}
B.~{Nelson}, E.~B. {Ford}, and M.~J. {Payne}.
\newblock {RUN DMC: An Efficient, Parallel Code for Analyzing Radial Velocity
  Observations Using N-body Integrations and Differential Evolution Markov
  Chain Monte Carlo}.
\newblock \emph{\apjs}, 210:\penalty0 11, January 2014.
\newblock \doi{10.1088/0067-0049/210/1/11}.

\bibitem[{Rajpaul} et~al.(2015){Rajpaul}, {Aigrain}, {Osborne}, {Reece}, and
  {Roberts}]{Rajpaul2015}
V.~{Rajpaul}, S.~{Aigrain}, M.~A. {Osborne}, S.~{Reece}, and S.~{Roberts}.
\newblock {A Gaussian process framework for modelling stellar activity signals
  in radial velocity data}.
\newblock \emph{\mnras}, 452:\penalty0 2269--2291, September 2015.
\newblock \doi{10.1093/mnras/stv1428}.

\bibitem[{Rajpaul} et~al.(2016){Rajpaul}, {Aigrain}, and
  {Roberts}]{Rajpaul2016}
V.~{Rajpaul}, S.~{Aigrain}, and S.~{Roberts}.
\newblock {Ghost in the time series: no planet for Alpha Cen B}.
\newblock \emph{\mnras}, 456:\penalty0 L6--L10, February 2016.
\newblock \doi{10.1093/mnrasl/slv164}.

\bibitem[Rasmussen and Williams(2005)]{Rasmussen2005}
Carl~Edward Rasmussen and Christopher K.~I. Williams.
\newblock \emph{Gaussian Processes for Machine Learning (Adaptive Computation
  and Machine Learning)}.
\newblock The MIT Press, 2005.
\newblock ISBN 026218253X.

\bibitem[{Reinhold} et~al.(2013){Reinhold}, {Reiners}, and
  {Basri}]{Reinhold2013}
T.~{Reinhold}, A.~{Reiners}, and G.~{Basri}.
\newblock {Rotation and differential rotation of active Kepler stars}.
\newblock \emph{\aap}, 560:\penalty0 A4, December 2013.
\newblock \doi{10.1051/0004-6361/201321970}.

\bibitem[{Rogers}(2015)]{Rogers2015}
L.~A. {Rogers}.
\newblock {Most 1.6 Earth-radius Planets are Not Rocky}.
\newblock \emph{\apj}, 801:\penalty0 41, March 2015.
\newblock \doi{10.1088/0004-637X/801/1/41}.

\bibitem[{Russell}(1906)]{Russell1906}
H.~N. {Russell}.
\newblock {On the light variations of asteroids and satellites}.
\newblock \emph{\apj}, 24:\penalty0 1--18, July 1906.
\newblock \doi{10.1086/141361}.

\bibitem[{Scargle}(1982)]{Scargle1982}
J.~D. {Scargle}.
\newblock {Studies in astronomical time series analysis. II - Statistical
  aspects of spectral analysis of unevenly spaced data}.
\newblock \emph{\apj}, 263:\penalty0 835--853, December 1982.
\newblock \doi{10.1086/160554}.

\bibitem[{Skumanich}(1972)]{Skumanich1972}
A.~{Skumanich}.
\newblock {Time Scales for CA II Emission Decay, Rotational Braking, and
  Lithium Depletion}.
\newblock \emph{\apj}, 171:\penalty0 565, February 1972.
\newblock \doi{10.1086/151310}.

\bibitem[Ter~Braak(2006)]{terBraak2006}
Cajo~JF Ter~Braak.
\newblock A markov chain monte carlo version of the genetic algorithm
  differential evolution: easy bayesian computing for real parameter spaces.
\newblock \emph{Statistics and Computing}, 16\penalty0 (3):\penalty0 239--249,
  2006.

\bibitem[ter Braak and Vrugt(2008)]{terBraak2008}
Cajo~JF ter Braak and Jasper~A Vrugt.
\newblock Differential evolution markov chain with snooker updater and fewer
  chains.
\newblock \emph{Statistics and Computing}, 18\penalty0 (4):\penalty0 435--446,
  2008.

\bibitem[{Vanderburg} et~al.(2015){Vanderburg}, {Montet}, {Johnson},
  {Buchhave}, {Zeng}, {Pepe}, {Collier Cameron}, {Latham}, {Molinari}, {Udry},
  {Lovis}, {Matthews}, {Cameron}, {Law}, {Bowler}, {Angus}, {Baranec},
  {Bieryla}, {Boschin}, {Charbonneau}, {Cosentino}, {Dumusque}, {Figueira},
  {Guenther}, {Harutyunyan}, {Hellier}, {Kuschnig}, {Lopez-Morales}, {Mayor},
  {Micela}, {Moffat}, {Pedani}, {Phillips}, {Piotto}, {Pollacco}, {Queloz},
  {Rice}, {Riddle}, {Rowe}, {Rucinski}, {Sasselov}, {S{\'e}gransan},
  {Sozzetti}, {Szentgyorgyi}, {Watson}, and {Weiss}]{Vanderburg2015}
A.~{Vanderburg}, B.~T. {Montet}, J.~A. {Johnson}, L.~A. {Buchhave}, L.~{Zeng},
  F.~{Pepe}, A.~{Collier Cameron}, D.~W. {Latham}, E.~{Molinari}, S.~{Udry},
  C.~{Lovis}, J.~M. {Matthews}, C.~{Cameron}, N.~{Law}, B.~P. {Bowler},
  R.~{Angus}, C.~{Baranec}, A.~{Bieryla}, W.~{Boschin}, D.~{Charbonneau},
  R.~{Cosentino}, X.~{Dumusque}, P.~{Figueira}, D.~B. {Guenther},
  A.~{Harutyunyan}, C.~{Hellier}, R.~{Kuschnig}, M.~{Lopez-Morales},
  M.~{Mayor}, G.~{Micela}, A.~F.~J. {Moffat}, M.~{Pedani}, D.~F. {Phillips},
  G.~{Piotto}, D.~{Pollacco}, D.~{Queloz}, K.~{Rice}, R.~{Riddle}, J.~F.
  {Rowe}, S.~M. {Rucinski}, D.~{Sasselov}, D.~{S{\'e}gransan}, A.~{Sozzetti},
  A.~{Szentgyorgyi}, C.~{Watson}, and W.~W. {Weiss}.
\newblock {Characterizing K2 Planet Discoveries: A Super-Earth Transiting the
  Bright K Dwarf HIP 116454}.
\newblock \emph{\apj}, 800:\penalty0 59, February 2015.
\newblock \doi{10.1088/0004-637X/800/1/59}.

\bibitem[{Wolfgang} et~al.(2015){Wolfgang}, {Rogers}, and {Ford}]{Wolfgang2015}
A.~{Wolfgang}, L.~A. {Rogers}, and E.~B. {Ford}.
\newblock {Probabilistic Mass-Radius Relationship for Sub-Neptune-Sized
  Planets}.
\newblock \emph{ArXiv e-prints}, April 2015.

\end{thebibliography}
\end{document}